\documentclass{jfm}

\usepackage{graphicx}
\usepackage{newtxtext}
\usepackage{physics}
\usepackage{newtxmath}
\usepackage{natbib}
\usepackage{subfigure}
\usepackage{hyperref}
\usepackage{lipsum}
\usepackage{tikz}
\usepackage{caption}
\hypersetup{
    colorlinks = true,
    urlcolor   = blue,
    citecolor  = black,
}

\newcommand{\RomanNumeralCaps}[1]
\linenumbers

\newcommand{\co}{\, \textit{,}}
\newcommand{\dt}{\, \textit{.}}

\title{Cross-flow oscillations of a circular cylinder with mechanically coupled rotation}

\author{A. Nitti \aff{1},
  G. De Cillis \aff{2}
 \and M. D. de Tullio \aff{1}\corresp{\email{marcodonato.detullio@poliba.it}}}

\affiliation{\aff{1}Department of Mechanics, Mathematics and Management, Polytechnic University of Bari, Via Re David 200, 70125 Bari, Italy \\
\aff{2}Euro-Mediterranean Centre on Climate Change Foundation, Ocean Predictions and Applications Division, Via Augusto Imperatore 16, 73100, Lecce, Italy}

\begin{document}
\maketitle

\begin{abstract}
Flow-Induced Vibrations (FIV) of an elastically mounted circular cylinder are investigated by means of two-dimensional simulations. A mechanical coupling between cross-flow translation and rotation provides a single degree-of-freedom system in which the coupled rotational oscillations affect the fluid-structure dynamics. The structural response of this system is investigated exploring the design space spanned by reduced velocity, coupling radius, and phase density ratio. The kinematic coupling introduces the rotation-induced shear layer modifications, as well as an equivalent inertia effect connected to the coupling force. Such a computational campaign is carried out by means of direct numerical simulations with immersed boundary forcing at a Reynolds number equal to 100. The investigated system exhibits the wake-body synchronisation features typical of lock-in for non-rotating cylinders. However, the kinematic coupling provides a novel FIV scenario, in which the oscillation amplitude is magnified in the locked configurations with respect to the forced rotation case. Furthermore, it is found a significant widening of the reduced velocity domain where the lock-in condition takes place. In view of the proposed analyses, it is determined that the coupled rotation guarantees the phase alignment between lift and displacement necessary to sustain the lock-in condition, making the oscillation amplitude grow indefinitely with the reduced velocity. This is inherently achieved due to the rotational shear layer and the added mass contribution, which prevent the exact match between oscillation frequency and system natural frequency in vacuum. The outcomes of this study might potentially lead to innovative water energy harvesters offering larger power outputs and extended optimal operating regions.
\end{abstract}



\section{Introduction} \label{sec:intro}
The elastic response of bluff bodies subject to cross-flow oscillations collects significant interest from the scientific community owing to its presence in a broad range of applications. The mutual energy exchange in Flow-Induced Vibrations (FIV) can compromise the integrity of immersed structures \citep{paidoussis1998fluid} or be harnessed to harvest renewable energy from clean sources \citep{seyed2020flow,young2014review}. In this connection, the study of FIV of elastically mounted circular cylinders allowed us to elucidate the fundamental mechanisms behind the related Fluid-Structure Interaction (FSI) dynamics \citep{bearman1984vortex,anagnostopoulos1992response,williamson2004vortex,sarpkaya2004critical,bearman2011circular}, as well as it provided a proof of concept for predicting the aforementioned engineering scenarios. \\
The elastically mounted circular cylinder has been the subject of numerous patents \citep{drew2009review} proposing water energy converters based on the enhancement and stabilisation of the cylinder VIV dynamics. In the renewable generators landscape, circular cylinders belong to the class of Alternating Lift Technologies (ALTs), typically targeted towards water flows slower than $1.0-1.5$ m/s, which cover a range of velocities inaccessible to watermills and turbines \citep{yuce2015hydrokinetic}. Considering that the majority of sea and river currents fall within this range, and that these currents are highly predictable and seasonally consistent compared with wind or waves \citep{bahaj2003fundamentals}, such converters can occupy a relevant part of the renewable energy portfolio. Compared with tidal turbines they need a much smaller intake volume, and extract more energy on a peer projected area basis, thus resulting in larger power-to-volume density and a reduced impact on marine life \citep{bernitsas2006vivace}. These motivations make the ALTs energy harvesters a current avenue of research and optimisation \citep{bernitsas2008vivace,hobbs2012tree,abdelkefi2012phenomena,wang2020state}, especially regarding oscillating cylinders. 
A successful implementation of the present concept was realised by \cite{kim2016performance}, who tested a 4 kW, four-cylinder prototype achieving the $88.6$ \% peak global efficiency of the Betz limit. Most of the marine turbines provide a $50.6$ \% conversion efficiency of the Betz limit at a nominal flow speed faster than $2.0$ m/s, and they require significantly larger initial capital cost \citep{wilberforce2019overview}. In 2016 the Technology Readiness Level (TRL) was assessed 7 out of 9 by the US Department of Energy \citep{kim2016performance,sun2020adaptive}, indicating an advanced level of maturity of this technology. To this extent, coupled computational investigations can provide a cheap design platform to dissect the VIV mechanisms and test design advancements. \\
The mechanical power $P$ potentially harvested from an oscillating cylinder with diameter $D$ and length $L$ is proportional to
\begin{equation}
P \propto \rho_f U^2 f_y A_y D L \co 
\end{equation}
with $A_y$ the oscillation amplitude, $f_y$ the oscillation frequency, $U$ the free-stream flow velocity, $\rho_f$ the fluid density. Thus, for a given oscillation frequency, a straightforward enhancement of the extracted power comes from the amplification of the amplitude of VIV oscillations. The present work takes steps to investigate an energetically efficient way to augment the power extracted from elastically mounted cylinders undergoing VIV. \\   
A simple two-dimensional (2D) cylinder free to oscillate in the cross-flow direction is known to experience VIV owing to the repeated shedding of vortices from alternating sides at a characteristic Strouhal frequency, which is a function of the Reynolds number \citep{chen1985flow}. When the vortex-shedding frequency is far from the natural frequency of the cylinder, the vibrations frequency matches the shedding frequency. However, as the shedding frequency approaches the natural frequency of the system in vacuum, the lock-in regime is established \citep{khalak1999motions}, and the former is determined by the latter. Under lock-in, the shedding frequency remains locked at the system natural frequency when increasing the hydrodynamic loading conditions, and the body exhibits Large Amplitude Oscillations (LAOs), until the wake becomes desynchronised, and the amplitude drops down to small values \citep{govardhan2000modes}. Under broken flow symmetry, elastically mounted bodies may also be subjected to the phenomenon of galloping \citep{paidoussis2010fluid}, which consist of an instability mechanism characterised by low-frequency oscillations, whose amplitude generally increases with the hydrodynamic loading. Galloping is driven by the asymmetric pressure distribution around the body, therefore, it does not necessarily involve a synchronisation with the vortex dynamics, although galloping and VIV regions might overlap \citep{corless1988model}. \\        
In this scenario, different passive and active strategies have been proposed to either amplify or suppress VIV, depending on the engineering context, whose overview can be found in \citep{kumar2008passive,wang2016active}. One of the most effective means of controlling the cross-flow cylinder oscillations relies on the enforcement of a prescribed rotatory motion to the cylinder itself. \citet{bourguet2014flow} explored the cross-flow vibrations of a 2D circular cylinder subjected to forced rotation at Reynolds number $\text{Re}=100$, illustrating that the peak oscillation amplitude can be even tripled with respect to the non-rotating cases. The rotation was observed to force the system to the lock-in condition up to a certain rotation rate, depleting the galloping regime (despite the rotation introducing a relevant asymmetry on pressure distribution). Results from this investigation are confirmed by experimental measurements \citep{seyed2015experimental} and further extended to larger rotation rates \citep{munir2021flow}. \\
The influence of rotary oscillations on elastically mounted cylinders have been first explored by \cite{du2015suppression}. Their study, aiming at suppressing VIV, showed by a numerical study at $\text{Re} \leq 400$ that forced rotary oscillations are able to suppress VIV amplitude to less than 1\% of the cylinder diameter. Furthermore, if the rotary speed rate is sufficiently high, the "lock-on" phenomenon occurs, where the shedding vortex frequency $f_s$ is determined by the forced rotational frequency $f_r$. This enables us to opportunely deviate the shedding frequency from the natural frequency of the spring-mass system $f_N$, thus preventing the onset of LAOs. However, despite being in a lock-on condition, for rotational frequencies close to the natural frequency, an amplification of oscillations with respect to the baseline non-rotating case is observed. An analogous system has been investigated experimentally by \citet{wong2018experimental} at $\text{Re}=2940$ covering a wider parameter space. They reported two lock-on regimes, related to the body oscillation frequency $f_y$, rather than the conventional vortex-shedding frequency. A rotary lock-on (RLO) regime is achieved when the body oscillation frequency settles on the forcing rotary oscillation frequency ($f_y\cong f_r$), whereas in the tertiary lock-on (TLO) regime, the body oscillation frequency is locked on the one-third subharmonic of the forcing frequency ($f_y\cong f_r/3$). The dynamic response showed similar trends with respect to the numerical results by \citet{du2015suppression}, despite differences in density ratio and Reynolds number. The cylinder vibration response has been investigated in the RLO regime with fixed forcing parameters, over a wide reduced velocity range. Three different regimes were reported, in two of which a galloping dynamics characterised by larger oscillations than the non-rotating case has been recognised. \\
In light of the beneficial effect of the rotatory motion for energy harvesting purposes, we present a novel mechanical system able to amplify the cylinder's oscillation amplitude by coupling cross-stream translation and rotation. This coupling can be achieved through a rack-and-pinion joint, thus, without further energy injection to sustain the cylinder rotation. When FIV occurs, this system produces an oscillatory-rotary motion which exhibits the typical features of lock-in over a broad parameter subset. A further beneficial outcome consists of the extension of the lock-in regime over a wider range of reduced velocities if compared with constrained rotation studies, with the consequent widening of the optimal operating condition limits. The FSI mechanisms of the system are investigated by means of two-dimensional Direct Numerical Simulation (DNS) relying on a well-established Immersed Boundary (IB) technique previously verified for different FSI problems \citep{de2016moving}, and results are outlined by means of a joint analysis of the body kinematics, vorticity patterns and hydrodynamic forces. \\
The parameter space includes the solid-to-fluid density ratio, the transmission coefficient, and the reduced velocity. It is worth pointing out that the present study provides a simple proof of concept of a new dynamical system, and practical considerations connected to its realisation, such as the transmission efficiency, or electrical generator coupling, are for the time being neglected. The suitability of the two dimensional approximation as well as the spanned non-dimensional parameters are discussed in details in the following section. \\
The manuscript is organised as follows. The physical model and the numerical method are presented in sections \ref{sec:model} and \ref{sec:num_met}, respectively. Then, the results of the computational campaign are assessed in section \ref{sec:results}. In particular, the structural response and the wake pattern are described in sub-sections \ref{ssec:struc} and \ref{ssec:wake}. The influence of fluid force distribution, including phase lag and added mass effects, are elucidated in sub-section \ref{ssec:forces}. Eventually, key findings and study outlooks are summarised in section \ref{sec:concl}. Additional information including validation and grid convergence studies are reported in the appendices.
\section{Computational model and numerical method} \label{sec:model_method}
\subsection{Model problem} \label{sec:model}
\begin{figure}
\subfigure[]{\includegraphics[width=0.49\columnwidth,trim={1.8cm 1cm 1.5cm 0},clip]{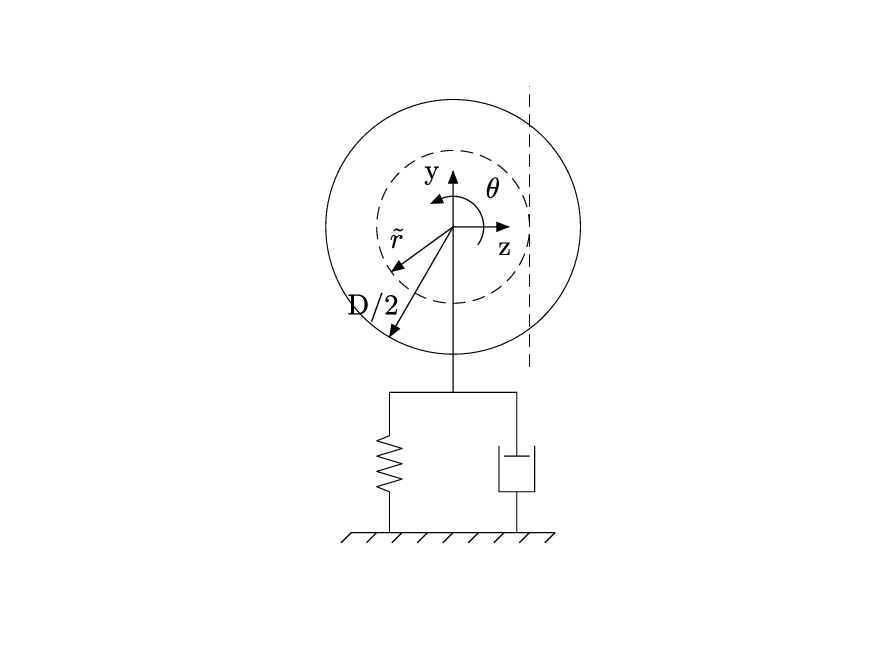}}
\subfigure[]{\includegraphics[width=0.49\columnwidth,trim={3.3cm 1.3cm 1.7cm 1.5cm},clip]{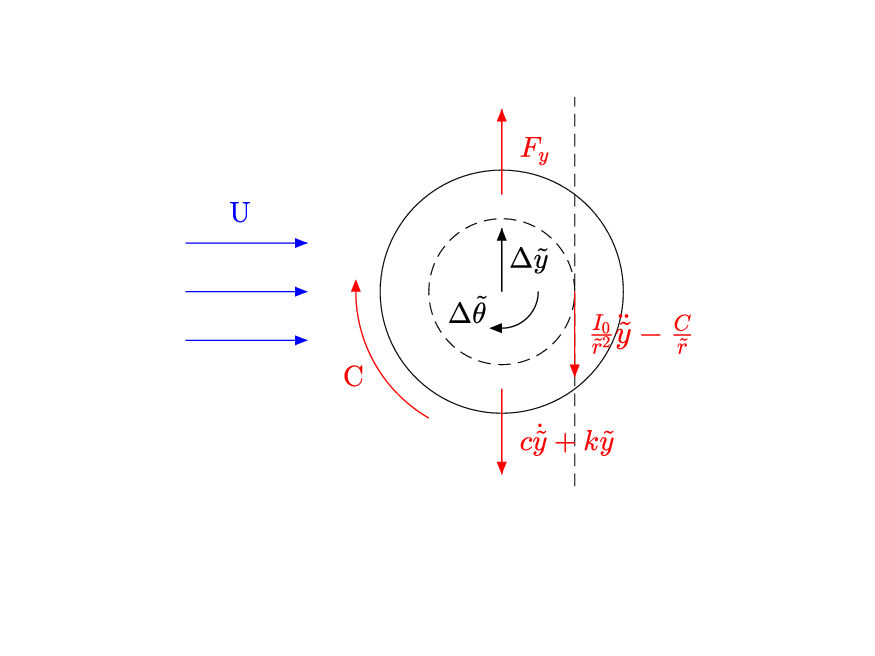}}
\caption{(a) Schematic of the computational domain and mechanical components. The vertical dashed line denotes the surface along which the pure rolling motion occurs. (b) Free body diagram. Red arrows represent the forces acting on the system undergoing a (positive) displacement depicted by black arrows.}
\label{fig:sch}
\end{figure}

The proposed model consists of a two-dimensional elastically mounted circular cylinder immersed in a cross-flow parallel to the $z$ axis. The cylinder can undergo both vertical and rotational displacement $(\tilde{y},\tilde{\theta})$, but the rotation is mechanically connected to the translation through a pure rolling motion with respect to a surface at distance $\tilde{r}$ from the cylinder centre. A sketch of the physical configuration is presented in figure \ref{fig:sch}(a). This results in a single degree-of-freedom system, governed by a forced second-order oscillator equation, which can be expressed as
\begin{equation}
\left(m+\frac{I_o}{\tilde{r}^2}\right) \ddot{\tilde{y}}+
c \dot{\tilde{y}}+ k \left(\tilde{y}-\tilde{y}_0\right)=F_y+\frac{C}{\tilde{r}} \dt
\label{eqn:dim_syst}
\end{equation}
The variables $\dot{\tilde{y}}$, $\ddot{\tilde{y}}$, $\dot{\tilde{\theta}}$ and $\ddot{\tilde{\theta}}$ represent the dimensional velocity and acceleration for the translational and rotational degrees of freedom, respectively. In the above equation $m$ and $I_o$ denote the cylinder mass and polar moment of inertia with respect to the cylinder centre; $c$ and $k$ are the damping and stiffness parameters, whereas $F_y$ and $C$ represent the resultant lift force and the torque exerted by the flow over the cylinder surface. Gravity is neglected in the present investigation. A free body diagram, from which the terms in equation \ref{eqn:dim_syst} can be inferred, is reported in figure \ref{fig:sch}(b). The equivalent non-dimensional formulation is obtained using the definitions $y=\tilde{y}/D$, $r=\tilde{r}/D$, $t=\tilde{t} \, U/D$, with $U$ being the free-stream velocity and $D$ the cylinder diameter. This yields the non-dimensional equation of rigid body motion,
\begin{equation}
\rho e \ddot{y}+ \mu\dot{y}+ \xi (y-y_0)= c_y+\frac{1}{r}c_m \co
\label{eqn:nondim_syst}
\end{equation}
which is characterised by the density ratio $\rho=\rho_s/\rho_f$, the damping parameter $\mu=c/\left(\rho_f U D \right)$, the non-dimensional stiffness $\, \xi=k/\left(\rho_f U^2\right)$ and the lift and torque coefficients $c_y=F_y/\left(\rho_f U^2 D\right)$, $c_m=C/\left(\rho_f U^2 D^2\right)$. The equivalent geometric parameter is defined as $e=A/D^2 + J_o/\left( D^2 \tilde{r}^2 \right)$, with $A$ the area of the cylinder and $J_o$ the polar moment of area. It addresses the cumulative inertial effect of the translating-rotating dynamics. The kinematic coupling is given by
\begin{equation}
r \dot{\theta}= \dot{y} \co
\label{eqn:kin_coulp}
\end{equation}
with $\dot{\theta}=\dot{\tilde{\theta}} D/U$ being the non-dimensional cylinder angular velocity. The relative loading condition of the dynamical system can be conveniently denoted in terms of reduced velocity $U_{r}=U/\left(f_n D\right)$, which represents the ratio of the natural time scale of the rigid body to the convective time scale of the flow, with $f_n=1/\left( 2 \pi \right) \sqrt{k /m_e}$ being the natural frequency of the oscillator in vacuum. The system damping is likewise characterised by the damping ratio $\zeta=c/\left(2 \sqrt{k m_e} \right)$ defined with respect to the critical damping. It is worth pointing out that, unlike previous investigations, the reduced velocity and damping ratio are defined with respect to the equivalent mass $m_e=m+I_o/\tilde{r}^2$, to account for the coupled kinematics.

The fluid motion is governed by the incompressible Navier--Stokes equations, which reads in non-dimendional form as
\begin{equation}
\begin{split}
& \nabla \cdot \mathbf{u}= 0 \co \\
& \pdv{\mathbf{u}}{t}+\left(\mathbf{u} \cdot \nabla \right) \mathbf{u}=
- \nabla p + \frac{1}{Re} \nabla^2 \mathbf{u} + \mathbf{f} \co
\end{split}
\label{eqn:gov_fluid}
\end{equation}
where $Re=U D/ \nu$ is the diameter-based Reynolds number, and $\mathbf{f}$ represents the body force arising from the IB treatment. All the simulations in the present study are conducted at a Reynolds number equal to $100$, in which the flow past an elastically mounted cylinder is acknowledged to be two-dimensional regardless of the other governing parameters, although laminar vortex shedding occurs. Along with the Reynolds number, the rotation rate plays a crucial role in the three-dimensional transition \citep{mittal2003flow,el2008three}. The numerical investigation by \cite{el2008three} pointed out that in rotating cylinders at $Re<500$ and $\dot{\theta}<1$ the rotation rate increases the critical Reynolds number for the three dimensionality, whereas at higher rotation rates the scenario becomes increasingly complex \citep{rao2013three}. \cite{munir2019numerical} identified the three-dimensional transition of a rotating cylinder at $Re=100$ for $\dot{\theta} = 7.6$. When the forced rotation is combined with VIV regimes, \cite{bourguet2014flow} found that at $Re=100$ a further delay in the three dimensional transition is observed over the whole parameter window where limit-cycle oscillations occur. At $(\dot{\theta},U_r)=(7.5,13.0)$ the flow was still found to be fully two dimensional. Following these guidelines we can expect the flow to be two dimensional within the explored parameter space. The latter was built spanning the reduced velocity $U_r$, the solid-to-fluid density ratio $\rho$ and the coupling coefficient $r$. The phase density ratio takes the values within the collection $\{4,6,8,10\}$, which includes realistic engineering scenarios for marine applications. We emphasise that, for any axisymmetric hollow cylinder, the moment of area and the density ratio represent a scale factor in the equivalent mass expression $m_e$, therefore, different density ratios might potentially correspond to different geometric configurations of the cylinder. The reduced velocity is sampled from $2.0$ to $12.0$ with a step of $0.4$, to capture the most relevant features at moderate loading conditions. Eventually, the transmission coefficient $r$ is varied from the initial value $r_0=0.5$ to lower values following the geometric progression $r_n=r_0 p^{n-1}$, since the equivalent area goes to infinity as it goes to zero with a hyperbolic trend. The common ratio is chosen to be $p=0.8$. Thus, $r$ takes the values $\{0.500,0.400,0.320,0.256,0.205,0.164,0.131,0.105,0.0839,0.0670\}$, where the lower bound provides doubled equivalent mass with respect to the purely translating case. In the space spanned by the prescribed $r$ and $U_r$ values very large oscillation amplitudes are encountered. Although the effect of the damping introduced by the user connection is of significant engineering interest, we set $\mu=0.0$ and postpone this issue to subsequent investigations to reduce the number of independent parameters in this preliminary study. All physical parameters involved in this study are summarised in table \ref{tab:params}. \\
Further insights on the proposed mechanical system can be found in Appendix \ref{sec:appB}, where the main construction aspects are briefly discussed.
\begin{table}
\centering
\begin{tabular}{lll}
Phase density ratio & $\rho$ & [4, 10]  \\
Coupling radius & $r$ & [0.067, 0.5]  \\
Reduced velocity & $U_r$ & [2.0, 12.0] \\
Damping ratio & $\zeta$ & 0.0 \\
Reynolds number & Re & 100 
\end{tabular}
\caption{Dimensionless physical parameters of the investigated system with the corresponding boundary values.}
\label{tab:params}
\end{table}
\subsection{Numerical method}
\label{sec:num_met}
The coupled fluid-structure system is simulated by means of the IB framework described by \cite{de2016moving} and \cite{nitti2020immersed}. The fluid phase equations \eqref{eqn:gov_fluid} are integrated by a classical fractional step method based on conservative second-order accurate centred differences over a staggered grid \citep{kim1985application}. Time integration is performed by a semi-implicit scheme, where the convective terms are advanced by an explicit third-order Runge-Kutta scheme and the diffusive terms by an implicit Crank-Nicholson scheme. The large penta-diagonal systems arising from the semi-implicit integration of the momentum equation are solved by means of an approximate factorisation, whereas the Poisson problem formulated to enforce continuity is solved with a direct solver to ensure mass conservation to a tight tolerance. Further details of the numerical method for the fluid phase can be found in \citep{orlandi2012fluid}. Following the approach presented in \citep{uhlmann2005immersed} the forcing term necessary to fulfil the no-slip condition is computed on Lagrangian markers laying on the immersed surface in the form of a volume force field, and then transferred to the Eulerian nodes. The information at the Lagrangian marker location is interpolated by means of a moving-least squares approach \citep{vanella2009moving}, which is recognised to attenuate spurious oscillations of hydrodynamic loads for moving interfaces while preserving second-order accuracy in space. The present solver has been validated and verified extensively for a broad variety of stationary and moving boundary problems in earlier works \citep{de2016moving}, as well as in the present study (see appendix \ref{sec:appA}). The rigid body equation of motion \eqref{eqn:dim_syst} is transformed into two first-order ordinary differential equations by a state-space formulation, which in turn are integrated by a fourth-order explicit Runge-Kutta scheme.  \\
The fluid and structural sub-systems are solved in a sequential fashion since the involved phase density ratios do not compromise the stability and the accuracy of the loosely coupled approach. The reader can find a numerical evidence for this assumption in the work by \cite{borazjani2008curvilinear}. The cylinder is immersed in a computational domain of $(z,y)$ dimensions $[-15D,25D]\times[-30D,30D]$, with an isotropic grid resolution in the subdomain $[-2D,8D]\times[-6D,6D]$ characterised by the spacing $\Delta x=0.015D$. A Dirichelet velocity boundary condition is used at the domain inlet, and a radiative outflow boundary condition \citep{orlanski1976simple} at the domain outlet. Free-slip condition are specified for the upper end lower boundaries. Thus, the baseline grid employed for the numerical campaign counts $801 \times 801$ nodes. Grid refinement studies are reported in appendix \ref{sec:appA} to confirm grid convergence. All simulations are performed with adaptive time-step size, adjusted to match the $\text{CFL}=0.2$ condition. The relative spacing between adjacent Lagrangian markers is set equal to $0.5 \Delta x$, with $\Delta x$ being the local Eulerian grid spacing, as numerical trade-off between accuracy of the interface condition and computational expense \citep{nitti2020immersed}. Given the present IB treatment, the flow field across the surface presents a smooth transition layer whose thickness takes at most two Eulerian cells,
as shown by \cite{de2016moving} with numerical experiments. Therefore, viscous and pressure loads are evaluated by interpolating the field variables at a probe created along the outward-pointing normal from the surface. The probe length is selected as the local averaged cell size. \\
All simulations are initialised with the periodic flow past a stationary cylinder at $Re=100$. The analyses are based on time series comprising the last 40 periods, collected after the initial transient dies out. Convergence of each simulation is established by monitoring the time-averaged and root-mean-square values of the fluid force coefficients and body displacement.
\section{Discussion of results} \label{sec:results}
\subsection{Structural response} \label{ssec:struc}
\begin{figure}
\subfigure[]{\includegraphics[trim=0.4cm 0 0 0, clip,width=0.49\columnwidth]{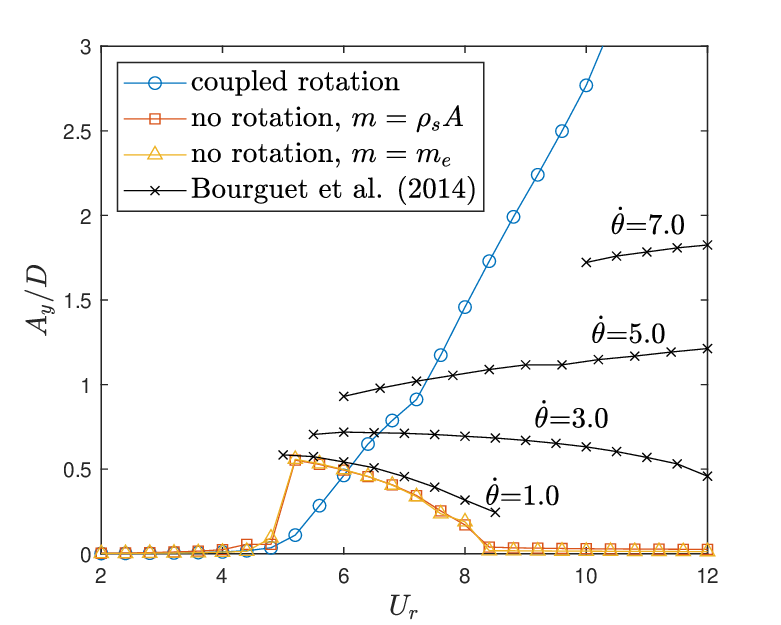}}
\subfigure[]{\includegraphics[trim=0 0 0.4cm 0, clip,width=0.49\columnwidth]{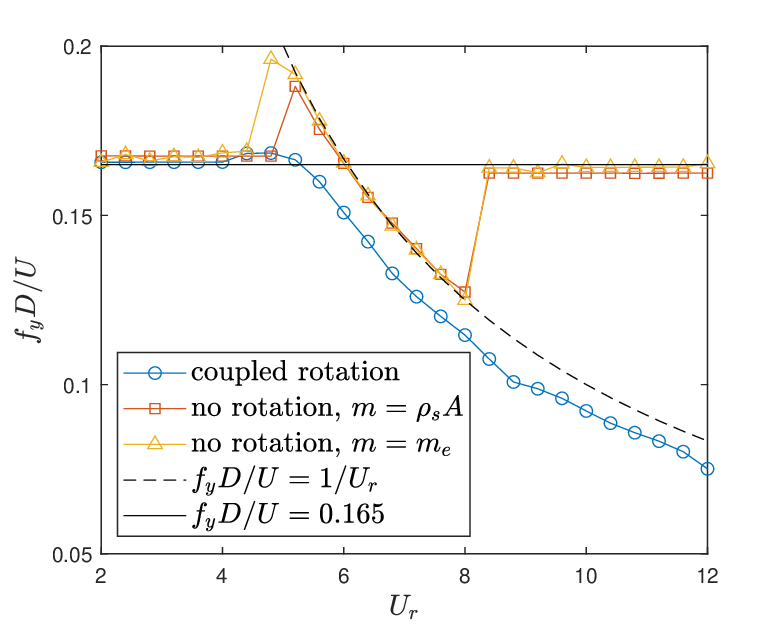}}
\caption{Maximum amplitude of vibration (a) and vibration frequency (b) as functions of the reduced velocity for cases with and without rotation coupling. Black lines in panel (a) represent the peak vibration amplitude of an elastically mounted cylinder with constant rotation rate under lock-in \citep{bourguet2014flow}.}
\label{fig:freq_disp_2}
\end{figure}

The structural response of the investigated model is quantified in the present section. In the first instance the displacement amplitude $A_y/D$ and the oscillation frequency $f_y D/U$ in the $\rho=8$ case are addressed. 
\begin{figure}
\subfigure[]{\includegraphics[trim=0.4cm 0 0 0, clip,width=0.49\columnwidth]{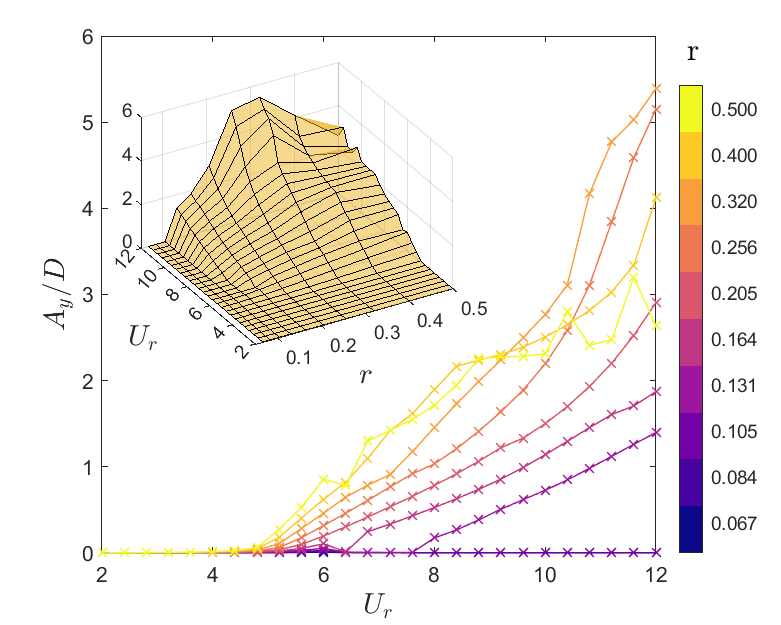}}
\subfigure[]{\includegraphics[trim=0 0 0.4cm 0, clip,width=0.49\columnwidth]{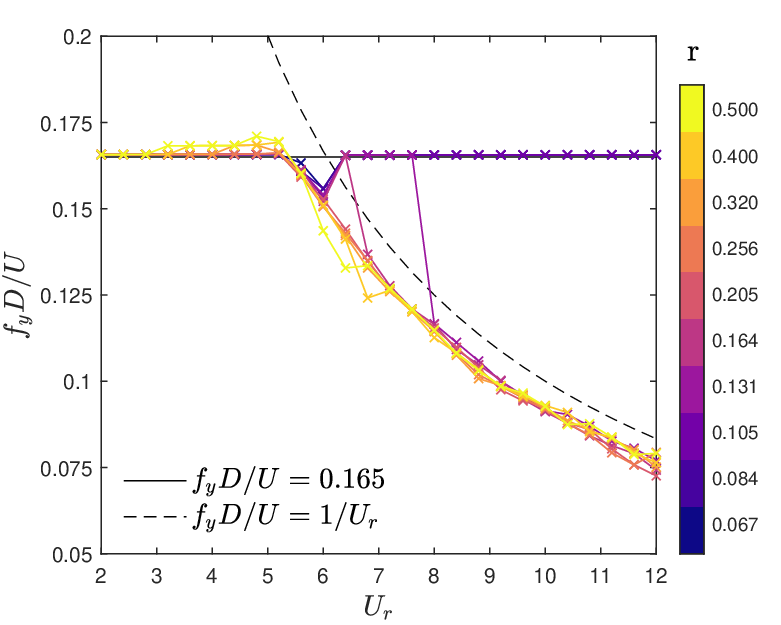}}
\caption{(a) Maximum oscillation amplitude and (b) vibration frequency as functions of the reduced velocity for different coupling radii. Each curve is associated with a value of coupling radius, consistently labelled in the adjacent colourbar. The insert in panel (a) provides a surface plot of the maximum oscillation amplitude for sake of clarity.}
\label{fig:freq_disp_1}
\end{figure}
\begin{figure}
\subfigure[$r=0.131$]{\includegraphics[width=0.49\columnwidth]{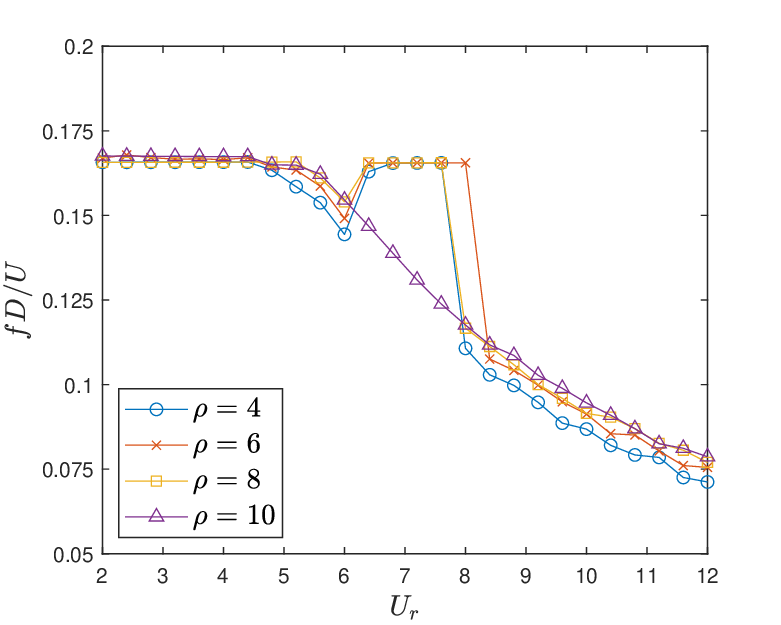}}
\subfigure[$r=0.164$]{\includegraphics[width=0.49\columnwidth]{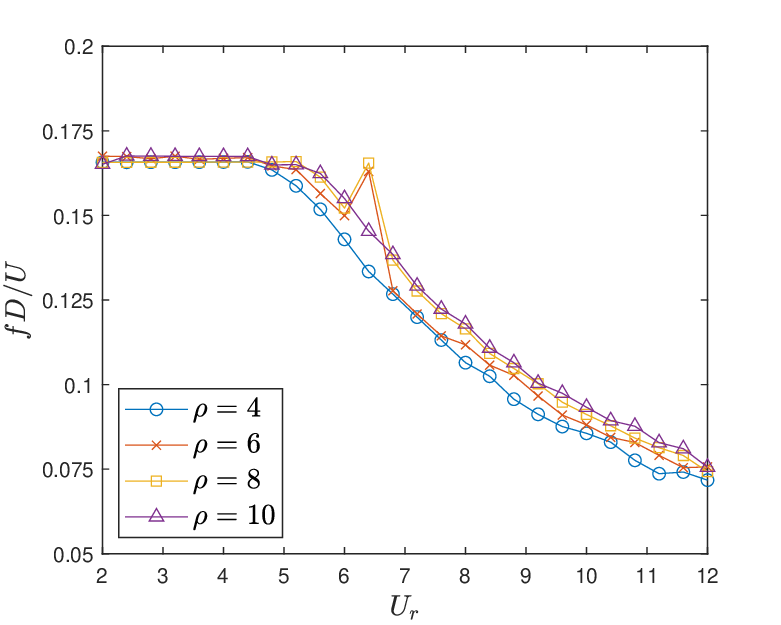}}
\caption{Dimensionless vibration frequency as a function of the reduced velocity for different phase density ratios. Two coupling radii are taken into account.}
\label{fig:rho_sweep}
\end{figure}
The vertical oscillations of the system are generally periodic and exhibit a strongly sinusoidal behaviour with null mean displacement. The LAO response is individuated over a wide range of parameters, demonstrating the effectiveness of the system as an energy harvester. The peak vertical displacement $A_y/D$ and oscillation frequency $f_y D/U$ corresponding to the maximum Power Spectral Density (PSD) are plotted as a function of the reduced velocity for different coupling radii in figure \ref{fig:freq_disp_1}. Any frequency value presented in the following of the manuscript is computed in the same way. As a general trend, the peak amplitude appears to grow monotonically when increasing the reduced velocity within the observed range. Conversely, the computational studies carried out by \cite{bourguet2014flow} and \cite{zhao2014vortex} showed that an elastically mounted cylinder with constant, uncoupled rotation undergoes LAOs in a narrower reduced velocity range on a peer rotational speed basis, and that an hysteretic behaviour takes place at the higher end of the lock-in region \citep{zhao2014vortex}. Reynolds number, density ratio and rotation rates are similar to those encountered in the present case. Such a feature enables our system as a truly versatile energy harvesting device, since a LAO dynamics is found for any hydrodynamic loading above the critical threshold, and no amplitude drop is found up to $U_r=12.0$. This is clearly evident from the comparison with computational data of elastic cylinders with constant rotation rate from \cite{bourguet2014flow}, available in figure \ref{fig:freq_disp_2} (a). The numerical study is limited to such a value because extremely large oscillations are already encountered. Larger oscillations are of limited practical interest owing to an increased design complication for a harvesting device. \\
Since the translational oscillations follows a sinusoidal kinematics, the peak rotational speed can be estimated as that of a purely alternating rolling motion by $\dot{\theta}_{\text{max}}=2 \pi f_y A_y /r$. Therefore, a picture of the maximum rotational speed can be simply drafted from figures \ref{fig:freq_disp_1}(a)-(b). The maximum rotational speed, equal to $\dot{\theta}=7.9$, is reached at $\left( r=0.320,U_r=12.0 \right)$. Relying on the outcomes of \cite{bourguet2014flow}, mentioned in section \ref{sec:model}, such a value does not entail the three-dimensional transition, therefore, 2D analyses are considered suitable. It is worth pointing out that the amplification effect generated by the enforced rotation can be already appreciated at $\dot{\theta}=1.0$, therefore, in the coupled model the FSI mechanisms are affected by rotation since the low end of the LAO region. The critical reduced velocity $U_r$ triggering the transition to the LAO regime decreases with the coupling ratio, showing a wide range of possibilities, from $U_r=7.6$ for $r=0.067$ to $U_r=4.8$ for $r=0.500$. \\
The oscillation frequency map (figure \ref{fig:freq_disp_1}(b)) exhibits two well-defined operating conditions. In the range of low reduced velocities the vibration frequency matches the vortex-shedding frequency observed in the static cylinder case, where $f_{\omega} D/U \in [0.163,0.167]$ \citep{zdravkovich1996different}. Under similar operating conditions a static cylinder immersed in a uniform flow generates an unsteady laminar wake pattern \citep{sumer2006hydrodynamics} inducing periodic loading oscillations. Analogous behaviour is observed at any reduced velocity for $r \leq 0.105$. Under this condition the cylinder frequency is synchronised with the predominant frequency of vortex shedding, and small amplitude vibrations ($A_y/D < 0.05$) take place. Conversely, each case undergoing LAO matches the natural frequency of the system $f_y D/U=1/U_r$ apart from a small offset. Thus, following the definition of \cite{khalak1999motions,williamson2004vortex} a lock-in condition is established. We emphasise that under lock-in the coupling radius does not affect the oscillation frequency, despite the fact that a broad range of peak rotation rates are encountered. From a purely structural perspective this feature is inherently achieved when considering the equivalent mass $m_e$ in the definition of reduced velocity. However, the mutual influence of hydrodynamic loading and rotation rate does not make this outcome foregone. Among the explored parameter combinations, the regime transition is found at $U_r=5.6$, except for $r=0.131$ and $r=0.164$, whose frequency jump transition at larger $U_r$ values. We emphasise that in the present model the LAO scenario could provide similarities with the RLO condition \citep{wong2018experimental}, where the cross-flow oscillation frequency settles on the frequency of forced rotary oscillations. This phenomenon was proved to generate an increase of the oscillation amplitude. Nevertheless, the kinematic coupling in the present model does not allow for a clear match of the RLO definition. Hence, the aforementioned regime will be simply referenced as lock-in hereafter. \\
At the largest coupling radii, the frequency correspondence does not trigger the LAO dynamics, but the transition is rather delayed towards larger values of reduced velocity. Whereas smaller coupling radii bring a larger equivalent inertia, it can be argued that the system is more resilient to the lock-in transition. However, this must not be correlated to the additional inertia, since the same phenomenon is observed in the same $U_r$ range at smaller phase density ratios $\rho$, but it is prevented at larger $\rho$. This can be definitely inferred from the frequency-reduced velocity plots for different $\rho$ and equal $r$ in figure \ref{fig:rho_sweep}. Hence, we speculate that the systems featured by a small coupling radius ($r=0.131$, $r=0.164$ with reference to figure \ref{fig:freq_disp_1} (b)) owe the delayed transition to a large resistant torque. In such configurations the rotation generates a viscous friction large enough to prevent the onset of LAO. This outcome highlights the complexity introduced by the coupled translations/rotational motion since \cite{wong2018experimental} found that, for enforced rotary oscillations characterised by a kinematic law with small amplitude and large frequency, the system is inclined to undergo LAs, under comparable reduced velocity values. \\
The variation of the coupling radius at constant reduced velocity provides a remarkable feature. For $U_r > 10.0$, the oscillation amplitude stops growing monotonically and reaches a maximum at $r=0.320$. This will be linked to the added mass effects in section \ref{ssec:forces}.  \\
The effect of the coupled kinematics on the structural response can be clarified by comparing amplitude and frequency data from the present model with non-rotating cases. Figure \ref{fig:freq_disp_2} provides a comparison of the $r=0.320$ case with two non-rotating configurations, one with identical mass and the other with $m=m_e$. It is immediately evident that the coupled kinematics generates a significant increase in the oscillation amplitude. Under forced rotary oscillations, \cite{wong2018experimental} showed a peak amplitude increase of approximately 270 \% at $U_r=8.0$, at large Reynolds numbers (Re $\approx 3000$). The present model likewise provides a 360 \% increase in the peak amplitude at the same reduced velocity. Unlike time-varying or constant forced rotation \citep{bourguet2014flow,zhao2014vortex,wong2018experimental}, our model seems to present no subcritical bifurcations (despite the fact that continuation algorithms are not employed), but a smooth transition to the LAO dynamics is observed. As previously documented, the non-rotating cases provide a sudden appearance of the LAO regime at $U_r=5.2$, with subsequent drop of the oscillation amplitude up to $U_r=8.4$ \citep{mittal2003flow}. Figure \ref{fig:freq_disp_2} (b) clearly reports the lack of a return to the static synchronisation regime (this refers to the synchronisation of the body vibration with the vortex frequency of a static cylinder) at $U_r \geq 8.8$, as well as the dependence of the frequency offset on the rotation. Following the arguments of \cite{bourguet2014flow}, this offset will be explained in section \ref{ssec:forces}, as a consequence of the added mass effect. Minor differences are noticed in the locked regime when comparing the non-rotating cases. Such a similarity takes place for $\rho = \mathcal{O}(10)$ or larger mass ratios since the extension of the lock-in region, as well as the amplitude of the corresponding oscillations, become nearly insensitive to the mass ratio itself \citep{khalak1997investigation,govardhan2004critical}.
\subsection{Wake pattern} \label{ssec:wake}
\begin{figure}
\centering
\subfigure[]{\includegraphics[width=0.49\columnwidth]{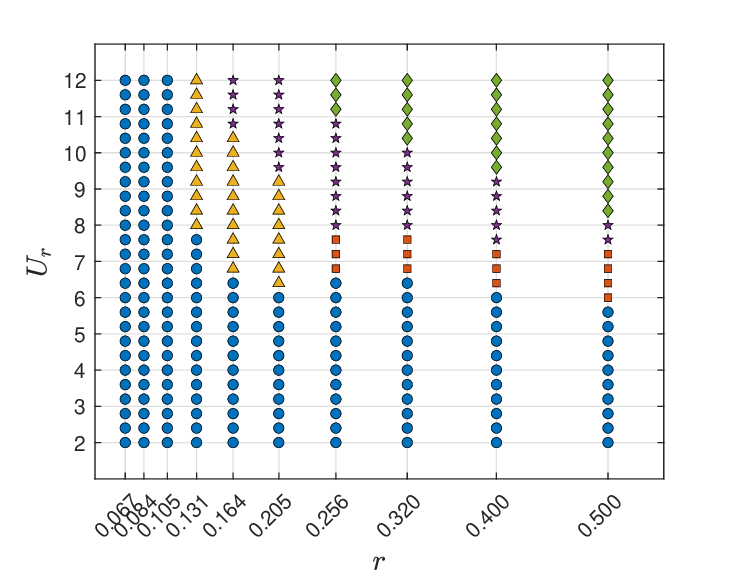}}
\subfigure[]{\includegraphics[width=0.49\columnwidth]{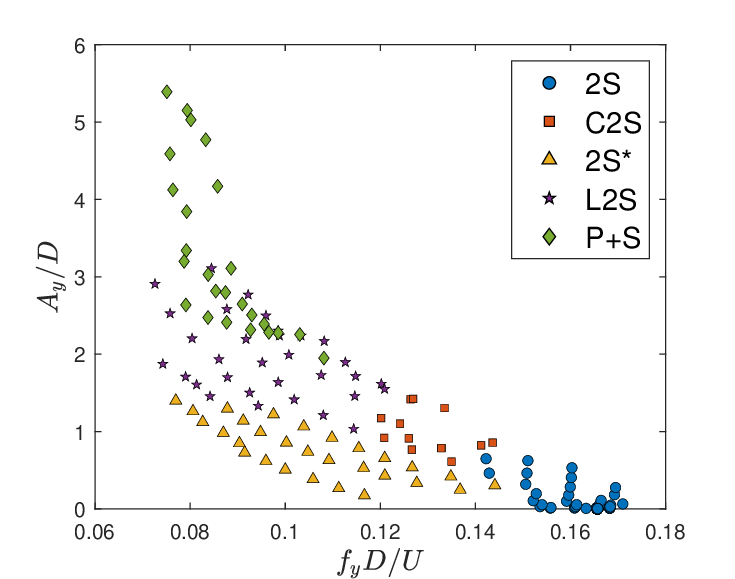}}
\subfigure[]{\includegraphics[width=0.49\columnwidth]{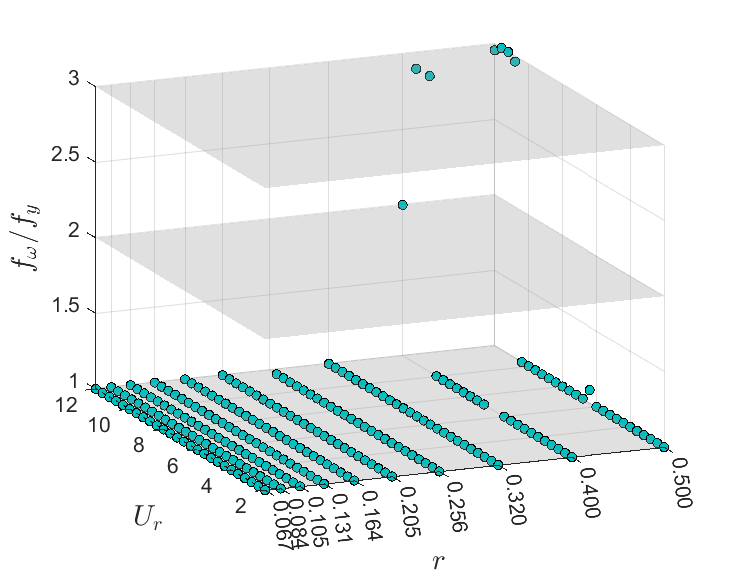}}
\caption{Wake pattern classification in the coupling radius-reduced velocity domain (a). Wake pattern classification in the peak oscillation amplitude-oscillation frequency domain (b). Wake frequency to oscillation frequency ratio and as a function of coupling radius and reduced velocity (c).}
\label{fig:wake_maps}
\end{figure}

\begin{figure}
\centering
\subfigure[2S pattern, $(r,U_r)=(0.5,3.2)$]{\includegraphics[trim=0.7cm 0 0.9cm 0, clip,width=0.49\columnwidth]{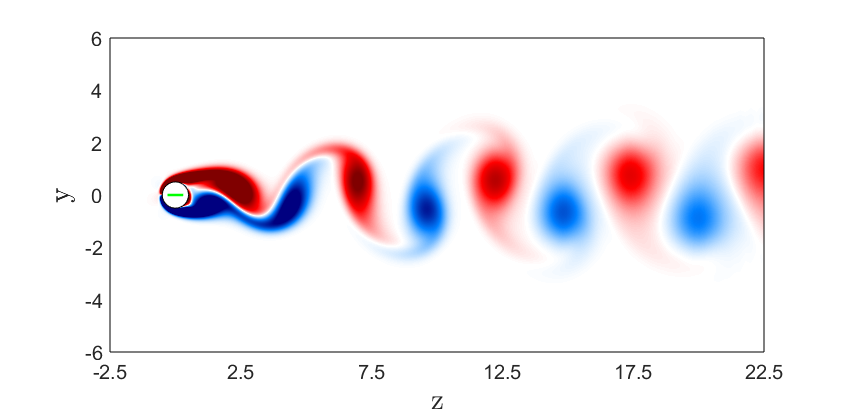}}
\subfigure[C2S pattern, $(r,U_r)=(0.4,6.8)$]{\includegraphics[trim=0.7cm 0 0.9cm 0, clip,width=0.49\columnwidth]{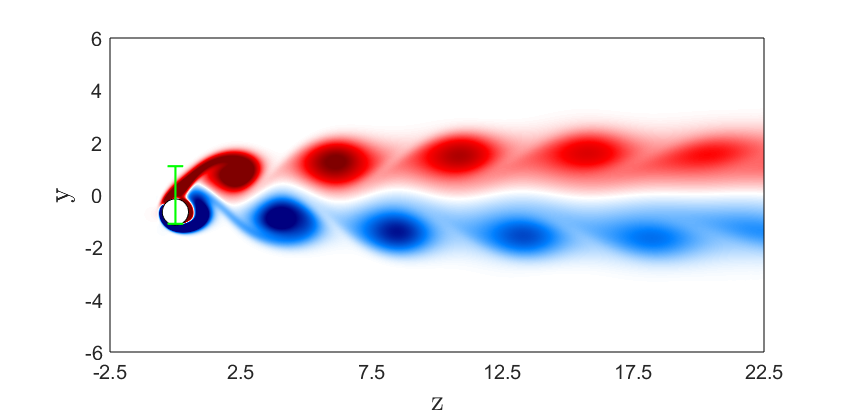}}
\subfigure[2S* pattern, $(r,U_r)=(0.164,7.6)$]{\includegraphics[trim=0.7cm 0 0.9cm 0, clip,width=0.49\columnwidth]{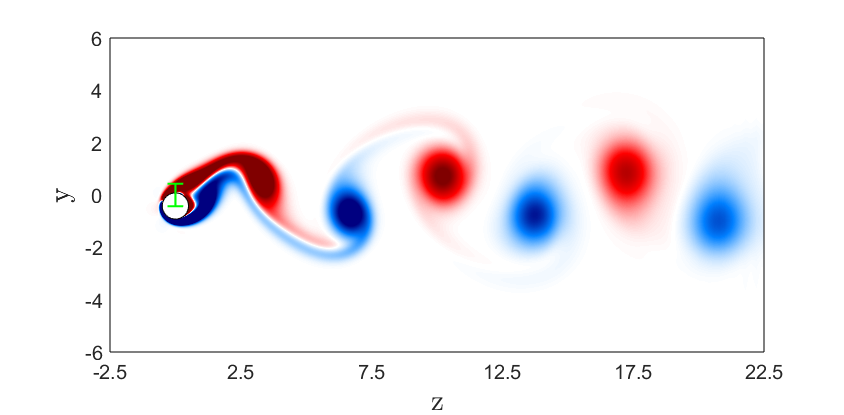}}
\subfigure[L2S pattern, $(r,U_r)=(0.256,9.6)$]{\includegraphics[trim=0.7cm 0 0.9cm 0, clip,width=0.49\columnwidth]{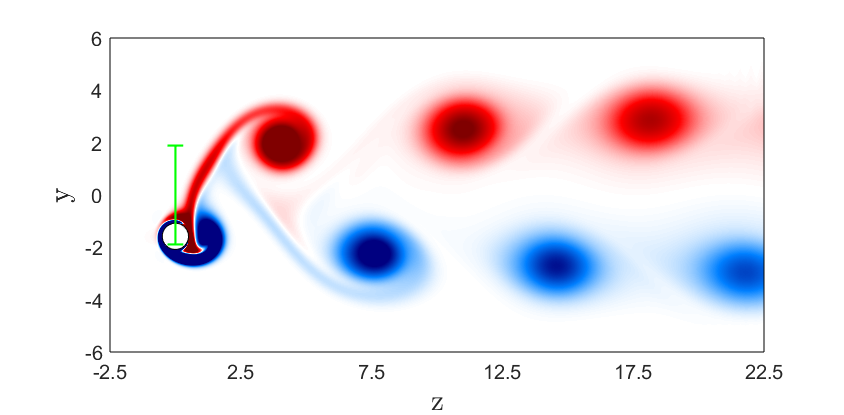}}
\subfigure[P+S pattern, $(r,U_r)=(0.4,11.2)$]{\includegraphics[trim=0.7cm 0 0.9cm 0, clip,width=0.49\columnwidth]{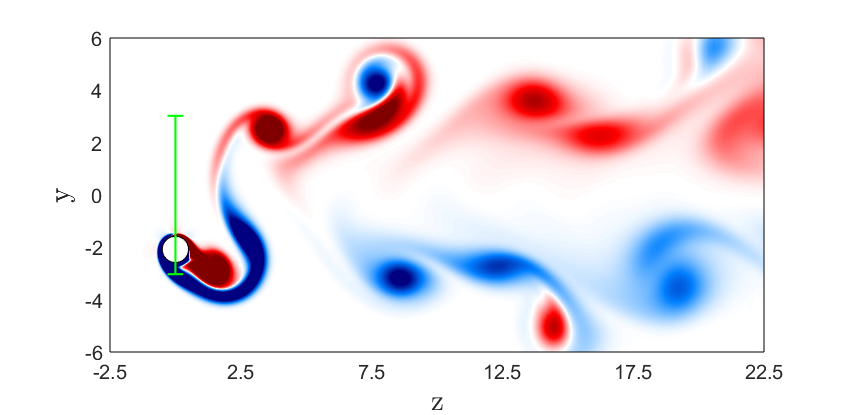}}
\caption{Instantaneous contours of spanwise vorticity identifying different wake patterns. The contour scale ranges from $\omega_x=-1.5D/U$ (dark blue) to $\omega_x=1.5D/U$ (dark red). The excursion covered by the cylinder is denoted by a cyan segment. A minor part of the computational domain is shown.}
\label{fig:vortcont1}
\end{figure}

The nature of vortex-shedding pattern downstream of a circular cylinder has been the subject of extensive investigations in previous studies \citep{griffin1976vortex,williamson1988vortex,tokumaru1991rotary,choi2002characteristics,govardhan2000modes,bourguet2014flow,wong2018experimental} due to its connection with the structural dynamics. Recently, \cite{menon2021initiation} quantitatively proved that the FIV of an elastically mounted cylinder is driven by the vorticity-induced force, and, specifically, the periodic oscillations are sustained by the shear layers on the transverse part of the cylinder surface, which in turn are energised by the wake structures. Thus, the inspection of the wake pattern is found to elucidate the influence of the vortex modes on the phase of lift and drag forces \citep{gabbai2005overview}. The referenced works have shown that both elastically mounted cylinders with rotary oscillation and constant rotation rate can exhibit a variety of recurring wake structures. Further interest concerning the wake pattern configuration arises when considering hydrokinetic energy converters built from multiple oscillating cylinders \citep{papaioannou2008effect,kim2016performance} in tandem/staggered configurations. In this context the wake tuning represents a crucial factor for maximising the energy harvesting potential. This section addresses the classification of the laminar wake patterns encountered in the present model, and it provides explanations of their link with the oscillation frequency. \\
A map of the different wake patterns for each pair $\left(r,U_r\right)$ investigated, is presented in figure \ref{fig:wake_maps} (a), where different patterns are classified by colours. Each pattern class, identified by the terminology of \cite{bourguet2014flow} and \cite{gabbai2005overview}, is illustrated in figure \ref{fig:vortcont1} by instantaneous contours of spanwise vorticity $\omega_x$, for certain values of $\left(r,U_r\right)$. A focus on the most common vortex-shedding modes can be found in \cite{gabbai2005overview}. In the present work the classification of the wake pattern is delivered by visual inspection, since transition phenomena might appear in the pattern switch zones \citep{prasanth2008vortex}. The interpretation of the wake pattern map can be enhanced by a collocation of each pattern in the frequency-amplitude space (figure \ref{fig:wake_maps} (b)). 
Both low hydrodynamic loading and low coupling radius regions are characterised by the 2S pattern, which is defined by two counter-rotating vortices shed per oscillation (figure \ref{fig:vortcont1}(a)), resembling the classic von Karman vortex street of a static cylinder at supercritical Reynolds number. At $r \leq 0.105$, the kinematic coupling provides a significant amplification of the rotational oscillations with respect to translation, but the amplitude of rotational oscillations is not large enough to modify the wake pattern. The C2S wake pattern is encountered within the frequency range $f_y D/U \in [0.12,0.14]$. Likewise, \cite{bourguet2014flow} found the C2S regime at frequencies lower than the Strouhal frequency, under a steady rotation regime. In the coupling radii region approximately delimited by ($0.256 \leq r \leq 0.5$) the vortices shed in the cylinder afterbody maintain a separate pattern: they tend to coalesce in the far wake (C2S) and to be advected downstream with limited rotational interaction. The rotational/translational acceleration of the cylinder at the peak displacement induces the rolling up of each of the separating shear layers into close vortices. The C2S pattern is found at moderate peak displacement (see figure \ref{fig:wake_maps}(b)). This trend was recognised for free vibrations of non-rotating cylinders \citep{singh2005vortex} and elastically mounted cylinders with steady forced rotation \citep{bourguet2014flow} in the range $4.8 \leq U_r \leq 6.0$. The same pattern is found here for larger reduced velocities ($6.0 \leq U_r \leq 8.0$). Interestingly, for $r \leq 0.205$, the 2S pattern is found under the lock-in regime as well, although vortices gain a larger spacing. This is mainly caused by the magnification of the vibration amplitude connected to the lock-in dynamics. The notation 2S* is formally introduced here to point out the relation between vortex-shedding pattern and dynamical regime. \\
\cite{wong2018experimental} denoted a 2S-P switching behaviour under forced rotational oscillations, outside the RLO region, owing to the inherently chaotic nature of the upper-branch VIV condition. A substantial analogy with the latter experimental study is found at very large peak oscillations ($A_y/D>1.5$), where two opposite-signed vortices of equal strength are alternatively shed, forming two parallel vortex rows with limited interaction. This pattern is here defined as L2S, and it was recognised by \cite{wong2018experimental} for $\dot{\theta}=4.0$. The related vortices have much higher enstrophy content than the 2S* and C2S cases, but they do not interact due to the lower shedding frequency, which increases their relative distance. This is confirmed by the clear separation of the C2S and L2S regions in the frequency domain shown in figure \ref{fig:wake_maps}(b). At larger oscillation amplitudes (thus, at higher rotation rates), the wake pattern is characterised by a blending of pairs (2P) and single vortices (S), therefore named P+S \citep{blackburn1999study,bourguet2014flow}. It is worth noting that each wake pattern, except the latter, covers a well-delimited region in the $(f_y D/U,A_y/D)$ domain. \\
Vorticity patterns characterised by two non-interacting vortices shed per cycle results in a more concentrated vorticity magnitude. This condition, which here takes place with the C2S and L2S patterns, was found to approximately coincide with the peak of lift forces seen in experimental results \citep{gabbai2005overview}, suggesting that larger forces are being induced by the shedding of more concentrated vorticity. As a matter of fact, figure \ref{fig:wake_maps}(b) correlates the C2S and L2S patterns with larger peak oscillation amplitude than the 2S pattern. \\
A further confirmation of the wake-body tuning is obtained by inspecting the maximum PSD frequency of the vorticity signal $f_{\omega}$. Vorticity time traces are collected by integrating the spanwise vorticity $\omega_x$ along the line $z/D=3.0$, in the cylinder wake. Such a measure collects the information on the flow structures being advected in the near wake of the cylinder, and it reflects the sinusoidal nature of the cylinder oscillation. The analysis of the wake-to-displacement frequency ratio $f_{\omega}/f_y$ reveals a clear tuning ($f_{\omega}/f_y \approx 1.0$) in the majority of the investigated cases. In fact, both the static synchronisation, and the lock-in regimes are characterised by a match of structural response and wake dynamics in the frequency domain (see figure \ref{fig:wake_maps} (c)). At $r \geq 0.4$ and $U_r \geq 10.0$, the dominant frequency of the vortex force jumps suddenly to a multiple of the displacement frequency: $f_{\omega}=2 f_y$ or $f_{\omega}=3 f_y$. In the referenced figure the desynchronised case lay over grey planes defining the integer multiples of the frequency ratio. It is achieved in correspondence of the P+S wake pattern, confirming that multiple vortex are shed per cycle without incurring in a chaotic vortex dynamics. This feature was recognised in the experimental investigation carried out by \cite{wong2018experimental} at much lower hydrodynamic loading and rotation rate ($U_r>5.5$ and $\dot{\theta} \geq 1.0$). \\
In the study carried out by \cite{wong2018experimental} and \cite{zhao2018experimental} the $f_{\omega}=3 f_y$ ratio is considered clearly indicative of the rotation-induced galloping response. This condition was found to be also characterised by a null lift-displacement phase difference (this feature will be addressed in the following paragraph) and by an unbounded growth of the oscillation amplitude with the reduced velocity. Although these conditions are verified in the present model, we also claim the correspondence of oscillation frequency and system natural frequency, which is typical of lock-in configurations. Hence, we account for a lock-in condition as the most general way to describe the system response across the reduced velocity domain.
\subsection{Force distribution} \label{ssec:forces}
\begin{figure}
\centering
\subfigure[$r=0.164$, $U_r=8.0$]{\includegraphics[width=0.8\columnwidth]{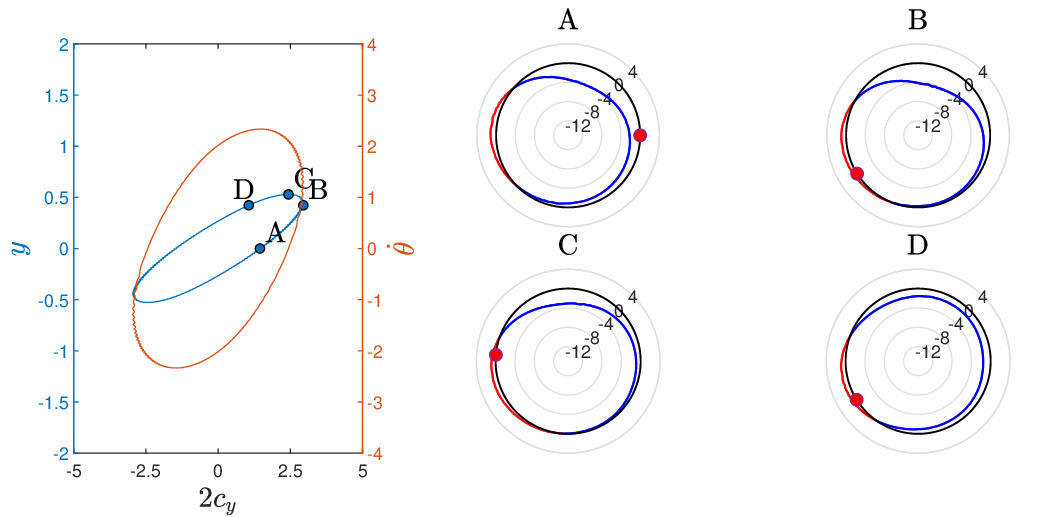}}
\subfigure[$r=0.320$, $U_r=8.0$]{\includegraphics[width=0.8\columnwidth]{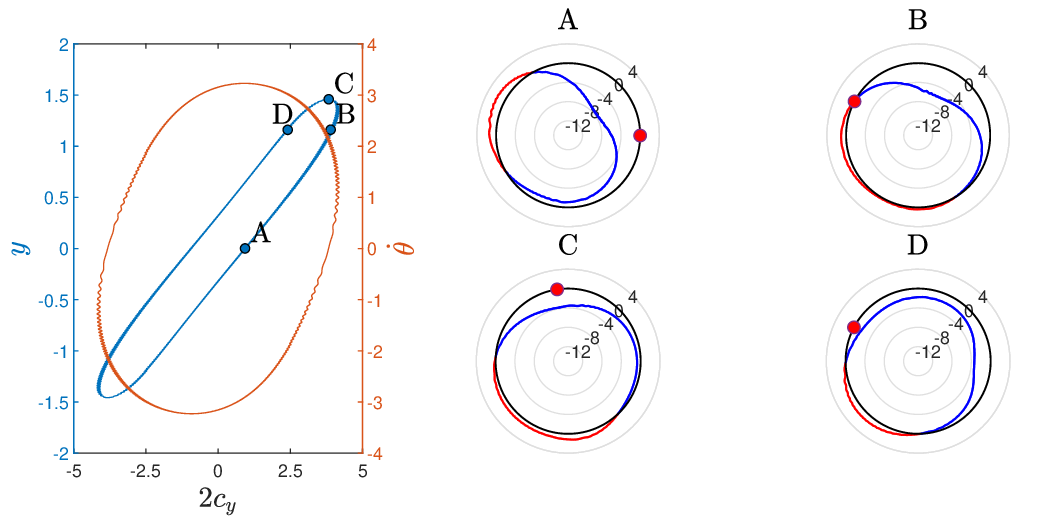}}
\subfigure[$r=0.500$, $U_r=8.0$]{\includegraphics[width=0.8\columnwidth]{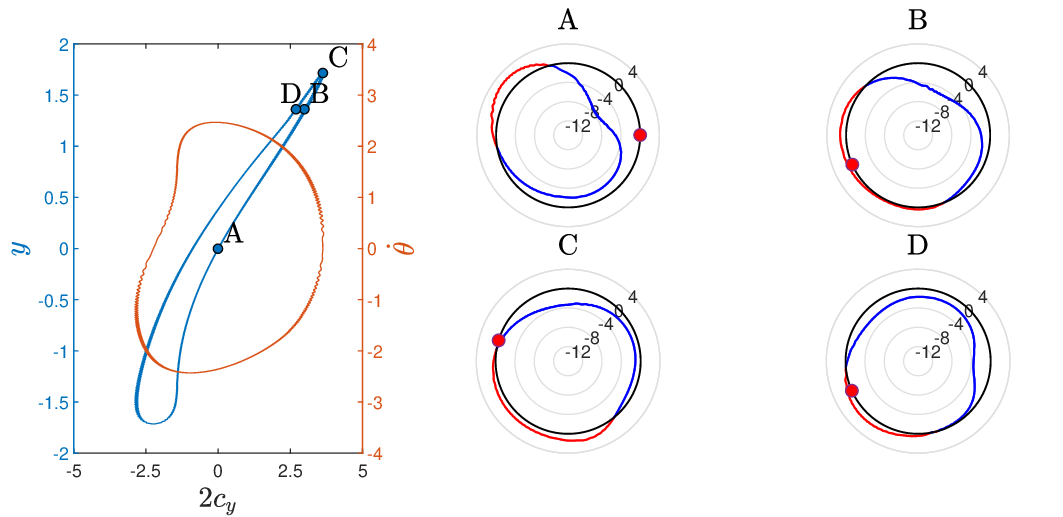}}
\caption{Phase portraits of cylinder displacement and rotational speed against cross-flow force coefficient (left panel). Instantaneous polar distribution of pressure coefficient $c_p=2( p-p_0 )/ ( \rho_f U^2 )$ in the time instants located over the phase portrait. Incoming flow impinges the cylinder on the left. The red dot denotes the instantaneous angular displacement of the cylinder.}
\label{fig:phaseport}
\end{figure}

\begin{figure}
\centering
\subfigure[]{\includegraphics[width=0.49\columnwidth]{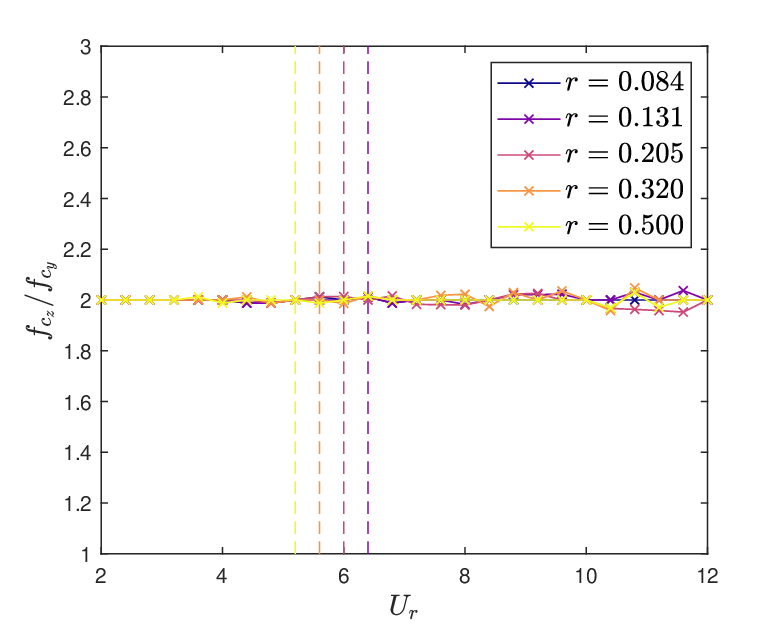}}
\subfigure[]{\includegraphics[width=0.49\columnwidth]{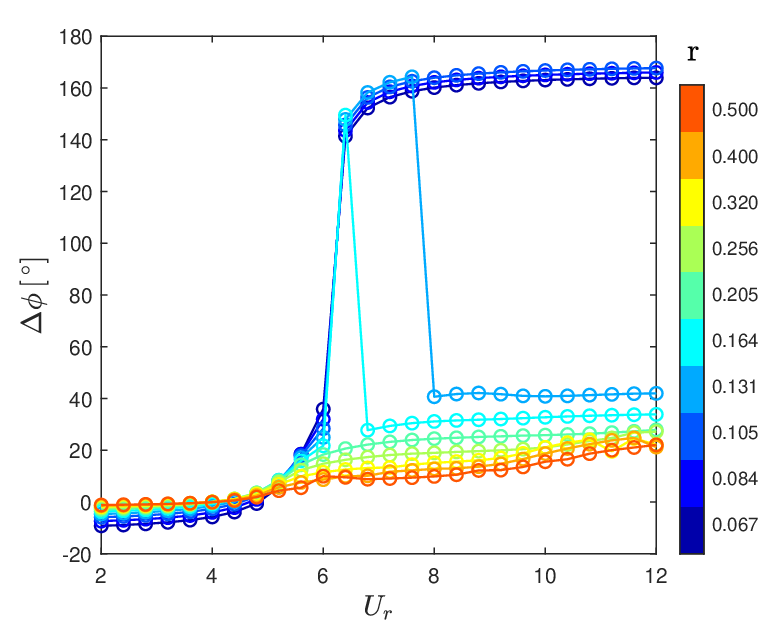}}
\caption{Dominant frequency ratio between cross-flow force coefficient $c_y$ and streamwise force coefficient $c_z$ against reduced velocity (b). Vertical lines represent the critical reduced velocity for transition to the LAO state. Phase difference $\Delta \phi$ between the cross-flow force coefficient $c_y$ and the cylinder displacement $y/D$ as a function of the reduced velocity for different coupling radii (b). Each curve is associated with a value of coupling radius, consistently labelled in the adjacent colourbar.}
\label{fig:ph_shift}
\end{figure}

Insights about the mechanisms behind the dynamics of the present oscillator are provided by inspecting features of fluid forces and energy transfer. \\
Phase portraits and instantaneous surface pressure are investigated for three cases sampled within the LAO region. For the same reduced velocity $U_r=8.0$, three coupling radii $r=\{0.164,0.320,0.500\}$ are examined. Despite being all within the lock-in region, they encompass significantly different oscillation amplitudes, as well as three different wake patterns: 2S* pattern for $r=0.164$, C2S pattern for $r=0.320$ and L2S pattern for $r=0.500$ (see figure \ref{fig:wake_maps} (a)). For each case, the phase portraits of vertical position and angular speed, both as a function of the lift (cross-flow) coefficient, are plotted in the left panels in figure \ref{fig:phaseport}. Phase portraits are depicted for the last five cycles of each simulation. Furthermore, the instantaneous pressure coefficient distribution is plotted in four relevant cycle instants, individuated as the null displacement instant during the upward swinging phase (A), the subsequent instant corresponding to the $0.8 A_y/D$ displacement value (B), the peak positive displacement instant (C), and the instant corresponding to $0.8 A_y/D$ displacement in the downward swinging phase (D). The pressure coefficient is defined over the Lagrangian markers as $c_p=2( p-p_0 )/ ( \rho_f U^2 )$. \\
The effect of the coupling radius can be noted from the superposed phase portraits when analysing the relative scaling between peak displacement and peak rotation rate. At $r=0.500$, for larger peak displacement, a lower peak rotation rate is achieved with respect to the case with $r=0.320$, at identical reduced velocity. Phase portraits corresponding to $r=0.164$ and $r=0.320$ prove the absence of higher harmonics in the lift force time history, and the steadiness of the phase difference between lift and displacement. It can be inferred that a unique limit cycle exists, and the oscillation frequency is locked onto the natural frequency of the system in vacuum instead of the Strouhal frequency of a fixed cylinder \citep{nobari2006numerical,placzek2009numerical}. The inclination of the lift-displacement cycle provides an estimate of the phase angle, therefore on the nature of the fluid-structure energy transfer. Such a feature will be explored for the whole configuration ensemble in the following. It is worth pointing out that the maximum rotational (therefore, translational) speed is achieved after the null-displacement point, halfway to the peak, due to inertial effects. For larger coupling radii (see, for instance, figure \ref{fig:phaseport} (b)), the peak speed is reached when the cylinder has a small net displacement, owing to a smaller equivalent inertia. From the phase portrait perspective this results in a larger inclination of the $\{c_y(t),y(t)\}$ plot. At instant A the cylinder is in the upward acceleration phase, mainly driven by the pressure unbalance caused by the negative pressure region on top. At the peak velocity (nearly point B), and peak displacement (point C), the suction area drifts counterclockwise on the cylinder surface, as well as the peak pressure spot. This is mainly connected with the detachment of a positive vortex. In position D the negative pressure area is evenly distributed on the surface and the cylinder motion is mainly governed by the elastic recoil. \\
For $r=0.320$, the cylinder experiences a larger peak oscillation amplitude, and a L2S wake pattern is established. In the central region of the phase portrait we observe a nearly linear variation of the displacement with the lift coefficient. Within this time window the inertial effects of the system are negligible due to a nearly constant velocity, and the fluid load distribution experiences small changes. Thus, the linearity of the elastic force allows for a nearly linear dynamics in between $y=-1$ and $y=1$. The system vibration frequency remains locked onto the natural frequency, and only two vortices of opposite sign are shed per cycle. The pressure snapshots provide a substantially different scenario with respect to the previous case. At null displacement (point A) both positive and negative pressure spots are biased towards the upper part of the cylinder surface, owing to the large rotation rate. \cite{mittal2003flow} observed a similar pressure coefficient distribution for a rotating cylinder at $Re=200$, confirming our speculations about the almost steady nature of hydrodynamic loads. In the other time instants the pressure forces follow the trend observed in the previous case, but with much larger negative pressure peaks.  \\
The wake pattern of the case $(r,U_r)=(0.500,8.0)$ has been classified as L2S (see figure \ref{fig:wake_maps}(b)), despite showing a transitional vortex dynamics. The corresponding phase portrait (figure \ref{fig:phaseport}(c)) therefore exhibits a well-defined path, but without the ovoid shape observed in the previous locked configurations. The symmetry in the $\{c_y(t),y(t)\}$ phase portrait is lost due to the appearance of secondary frequencies in the force coefficients. Comparing the phase portrait with the frequency plot (figure \ref{fig:wake_maps} (c)), we can speculate that, as pointed out by \citep{placzek2009numerical}, the wake is locked, since the main wake frequency matches the natural frequency in vacuum, and the additional high frequencies do not affect the cycle-to-cycle periodicity. The appearance of additional harmonics in the lift evolution can be related to the emission of the third vortex in the upper side of the wake, leading the transition to the $P+S$ wake regime. In the case under examination, the pressure distribution shows few changes with respect to the case $(r,U_r)=(0.320,8.0)$. Considering the similarity in amplitude and oscillation frequency, this further confirms that the body excitation is mostly sensitive to the pressure part of the force, as previously noted by \cite{bourguet2014flow} in the case of constant rotation rate.  \\
As recognised in earlier studies for static cylinders and forced cross-flow oscillations \citep{bishop1964lift}, the anti-symmetric nature of vortex shedding results in a ratio of 2 between the fundamental frequencies of the in-line and cross- flow force coefficients. For purely rotating cylinders, the symmetry breaking induced at large rotation rates ($\dot{\theta} \geq 3.5$) causes a switch to a frequency ratio $f_{c_z}/f_{c_y} \approx 1.0$ \citep{mittal2003flow}, i.e. matching the fundamental frequencies of $c_z$ and $c_y$. The same phenomenon was observed for elastically mounted cylinders with forced rotation outside the lock-in region \citep{bourguet2014flow}. In the present model the frequency match is found anywhere within the spanned parameter space, regardless of the cylinder dynamics (see figure \ref{fig:ph_shift} (a)). A slight misalignment is observed at large reduced velocities due to the emergence of secondary harmonics in the lift forces linked to the occurrence of the $P+S$ wake pattern. Similarly, \cite{bourguet2014flow} recognised significant PSD peaks for secondary and tertiary harmonics at $\dot{\theta} = 3.0$. \\
The phase difference $\Delta \phi$ between the cross-flow displacement $y(t)$ and the lift coefficient $c_y(t)$ can be interpreted as a descriptor of the transfer of mechanical energy (product of fluctuating lift force and centre of mass displacement) from the fluid to the cylinder \citep{gabbai2005overview}. In forced vibration studies the phase shift was associated with a change in the direction of energy transfer \citep{blackburn1999study,carberry2001forces,guilmineau2002numerical}, whereas a more complex scenario was evidenced for free vibrations. Experiments at high Reynolds number carried out by \cite{zdravkovich1982modification,khalak1999motions} showed that the phase jump takes place in the middle of the lock-in region, accompanied by a shift from the upper branch to the lower branch in the oscillation dynamics. Thus, the transition from the static synchronisation to the lock-in regime  does not entail a jump in the phase difference $\phi(c_y(t))-\phi(y(t))$. For low Re, no upper branch has been observed, therefore, the phase jump does not generate any discontinuous trend within the lock-in region. Specifically, \cite{prasanth2008vortex} detected a phase jump at $U_r=6.9$ in between of the LAO range ($5.0 \leq U_r \leq 8.3$), with the phase difference switching impulsively from $0^\circ$ to approximately $180^\circ$. The sudden phase shift was then associated to a change of the system response to small phase perturbations. Moreover, they pointed out that the phase jump is not hysteretic; the same behaviour is observed for increasing as well as decreasing $U_r$. Both computational and experimental studies \citep{prasanth2008vortex,govardhan2000modes} showed that the phase jump takes place when the oscillation frequency precisely matches the natural frequency of the structure in vacuum. This correspondence does not occur at the triggering of the lock-in phenomenon, where a small departure from the natural frequency is still verified due to fluid-structure feedback mechanisms \citep{prasanth2008vortex}, but in the middle of the lock-in region. This feature was found also in the present investigation for non-rotating cylinders, where the exact frequency match is achieved at $U_r=6.8$ (see figure \ref{fig:freq_disp_2} (b)). \\
In a similar fashion, for elastically mounted cylinders with forced rotation rate, both phasing states are observed in the lock-in region when $\dot{\theta} \leq 3$, although lower maximum phase differences are achieved \citep{bourguet2014flow,zhao2014vortex}. For larger rotation rates, force and displacement remain in phase, regardless of the hydrodynamic loading. In the former condition, an increase in the rotation rate corresponds to a reduction in the phase difference at large reduced velocities. Again, the phase jump was found to fulfil the condition of exact match between oscillation frequency and system natural frequency in vacuum. Both in rotating and non-rotating cylinders the phase difference was found to be affected mainly by the pressure part of the fluid force, especially in the range of low reduced velocity \citep{prasanth2008vortex,bourguet2014flow}. In forced vibration experiments \citep{williamson1988vortex} the jump in the phase difference has been attributed to a sharp change in the timing of vortex shedding when switching from the wake pattern 2S to pattern P, whereas for free oscillations no similar correspondences have been found. \\
In the present work the angular phase difference $[^\circ]$ measurement is based on discrete Fourier transform and maximum likelihood estimation of the signals’ initial phases \citep{sedlacek2005digital}. Figure \ref{fig:ph_shift} (b) provides a picture of the phase shift as a function of the reduced velocity for different coupling radii. For $U_r \leq 5.2$, the lift force is essentially in phase with the displacement, regardless of $U_r$ and $r$ values. For reduced velocities larger than $6.0$, the phase jump takes place only for the smallest values of the coupling radius, i.e. $r \leq 0.105$ (they do not undergo lock-in), reaching a peak difference of approximately $\Delta \phi \approx 180^\circ$. Interestingly, within the LAO case ensemble, the phase difference is found to gradually grow without a jump, but with a maximum phase difference inversely proportional to the coupling radius. The peak $\Delta \phi$ achieved for a LAO case is approximately $45^\circ$. Such proportionality owes to the larger peak rotation rate occurring with the smaller $r$ (see eq. \eqref{eqn:kin_coulp}). Likewise, recent experimental and numerical investigations \citep{seyed2015experimental,bourguet2014flow} showed that at larger rotation rates $\dot{\theta} > 3.0$ the phase difference jump vanishes in favour of a smooth transition. \\
We emphasise once more that, as long as the oscillation frequency of the system does not exactly match the frequency of the system in vacuum, the phase difference jump is prevented. The present system, once shifted to the lock-in regions, provides an almost constant frequency offset with respect to the natural frequency (see figure \ref{fig:freq_disp_1} (b)). Such a feature, along with the rotational motion, seems to guarantee the force-displacement phase alignment. A fundamental implication of this trend is the lack of a limiting mechanism for the oscillation amplitude of locked cases. \cite{prasanth2008vortex} verified the consequences of the phase jump by manipulating the phase shift in a LAO case under periodic motion. In such a case they found that forcing $\Delta \phi (t_0)=0$ generates very large amplitude oscillations. Our system is able to force this condition by a simple kinematic coupling, which results in magnified oscillations, as reported with the comparison in figure \ref{fig:freq_disp_2} (a). \\
As stated earlier, the described phenomenon is strictly connected with the offset of the oscillation frequency from the natural frequency, which in turn can be linked to the added mass effect. The natural frequency of an elastically mounted rigid body, immersed in a viscous fluid, differs from that of a dry body by a coefficient $C_a$ representing the additional inertia of the surrounding fluid displaced by the body itself. The mutual dependence of added mass, reduced velocity and natural frequency \citep{gabbai2005overview} makes the estimation of the correlation between added mass and hydrodynamic loading a cumbersome task. Following the discussion in \cite{vikestad2000added}, the "true" natural frequency $f_{nc}$, augmented by the added mass coefficient $C_a$, is defined by
\begin{equation}
f_{nc} = \frac{1}{2 \pi} \sqrt{\frac{k}{m_e+\rho_f A C_a}} \co
\label{eq:f_nc}
\end{equation}
where $\rho_f A$ is the fluid mass displaced by the body. Without any assumption on the harmonics content of the body kinematics, the effective added mass coefficient can be found from the lift coefficient and non-dimensional acceleration by \citep{vikestad2000added,bourguet2014flow,wang2021illuminating},
\begin{equation}
\displaystyle
C_a= -\frac{2}{\pi} \frac{\displaystyle \int_T c_y(t) \ddot{y}(t) \, dt}{\displaystyle\int_{T} \left( \ddot{y}(t) \right)^2 dt} \co
\label{eq:Ca}
\end{equation}
with the lift coefficient $c_y$ integrated in phase with the body acceleration. From equation \eqref{eq:f_nc}, the normalised oscillation frequency results in
\begin{equation}
\frac{f_y}{f_{nc}} = \sqrt{\frac{\rho e}{\rho e + \frac{\pi}{4} C_a}} \dt
\label{eq:fratio}
\end{equation}
Extensive experimental investigations \citep{khalak1997investigation,khalak1999motions,vikestad2000added}, over a broad range of Reynolds numbers, proved that the added mass coefficient decreases monotonically with the reduced velocity. Such a feature, as well as the existence of a negative added inertia region, has been generally found in vibrations of elastically mounted bluff bodies \citep{paidoussis2010fluid}. According to its definition, negative values of $C_a$ entail a change in the phase between fluid forces and body acceleration. When considering a flow past a cylinder with its own complex dynamics, there is no a priori explanation of the force-to-acceleration phase difference. To this extent, the evolution of the added mass with the reduced velocity (or with the natural frequency) has a rather complex interpretation. \cite{paidoussis2010fluid} proved with a simple linearized model of an elastically mounted bluff body that the added mass coefficient decreases monotonically with the reduced velocity as a result of the coupled dynamics of a wake oscillator and a solid oscillator. As a consequence, the true natural frequency $f_{nc}$ increases with the reduced velocity, as reported by numerous works \citep{gabbai2005overview}. \\
\begin{figure}
\centering
\subfigure[]{\includegraphics[width=0.49\columnwidth]{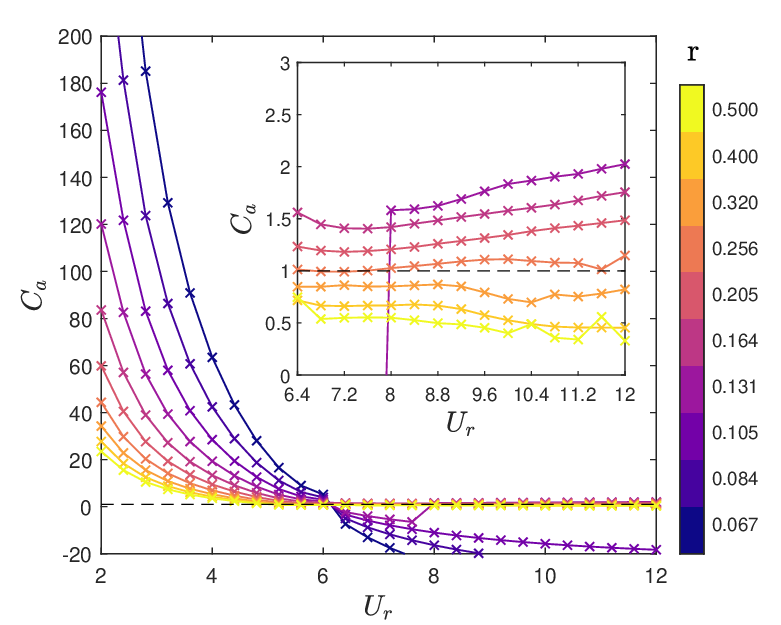}}
\subfigure[]{\includegraphics[width=0.49\columnwidth]{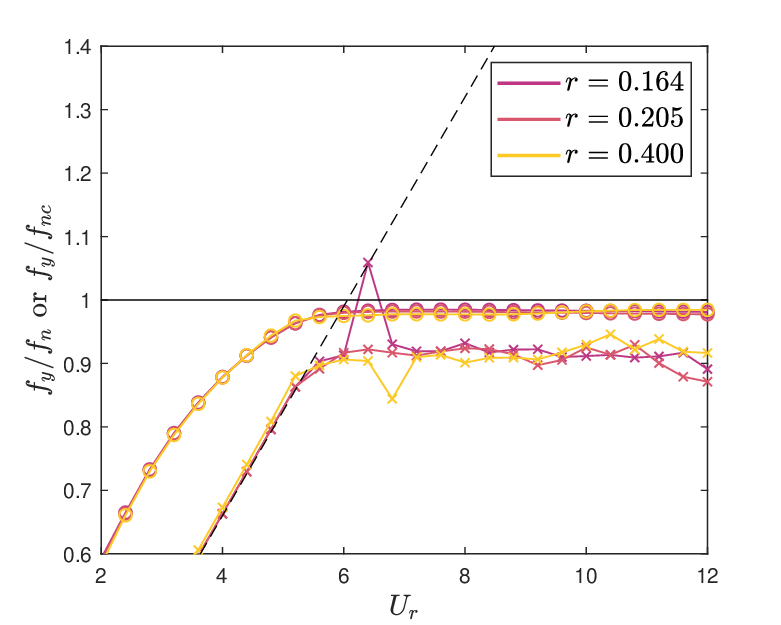}}
\subfigure[]{\includegraphics[width=0.49\columnwidth]{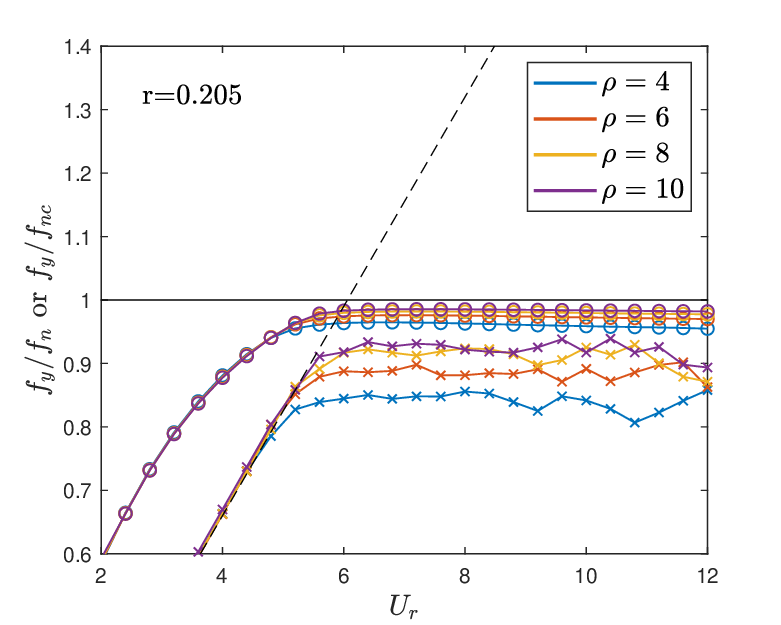}}
\caption{Effective added mass coefficient computed from expression \eqref{eq:Ca} as a function of the reduced velocity (a). Each curve is associated with a value of coupling radius, consistently labelled in the adjacent colourbar. Normalised frequency response as a function of the reduced velocity (b). Crosses in panel (b) indicate the $f_y/f_n$ ratio and circles indicate the $f_y/f_{nc}$ ratio, whereas the dashed line in the panel indicates the normalised frequency of vortex-shedding in the stationary cylinder case. Normalised frequency response as a function of the reduced velocity for the $r=0.205$ case, at different phase density values(c)}
\label{fig:amass}
\end{figure}
Figure \ref{fig:amass} shows the relation of the added mass coefficient $C_a$ and normalised frequency ratios ($f_y/f_{nc}$ or $f_y/f_n$) as a function of the reduced velocity, at different coupling radii. The added mass coefficient is found to decrease monotonically up to $U_r=6.4$, where it changes sign only for high equivalent mass cases ($r \leq 0.105$). The well-established link between added mass and lift force-displacement phase opposition \citep{paidoussis2010fluid} is found in this system too, by comparing the pairs $(r,U_r)$ entailing a phase opposition (figure \ref{fig:ph_shift} (b)) and a negative added mass coefficient (figure \ref{fig:amass} (b)). As a general trend, under lock-in the absolute value of $C_a$ decreases with the coupling radius, regardless of the oscillation amplitude, owing to the fact that a larger coupling ratio generally results in a smaller peak rotation rate. In this connection the growth of the rotation rate was found to be correlated with an increase of the added mass coefficient both at low Reynolds \citep{bourguet2014flow,zhao2014vortex} and high Reynolds numbers \citep{seyed2015experimental}. It can be inferred that a faster rotation generates a larger momentum diffusion in the surrounding fluid, therefore, a larger displaced fluid mass. A key difference with the numerical investigations with steady forced rotation rate \citep{bourguet2014flow,zhao2014vortex} relies in the $C_a$ behaviour within the early LAO region $6.4 \leq U_r \leq 12.0$. They highlighted a monotonic decrease of the added mass coefficient that, for small rotation rates, $\dot{\theta} \leq 3.5$ reaches negative values. Conversely, in the present work a non-classical trend is observed (see the insert in figure \ref{fig:amass}), with $C_a$ having a much moderate slope, despite the (positive) values interval being fully consistent. On the other hand, the same reference investigations recognised a similar trend for large rotation rates ($\dot{\theta} > 5.0$) and larger reduced velocity ($U_r>14.0$). One can notice that the oscillation amplitude is found to be maximum in correspondence of a nearly unitary added mass coefficient, by comparing figure \ref{fig:amass} (a) and figure \ref{fig:freq_disp_1} (a). The unity line corresponds to the added mass coefficient for a non-rotating cylinder based on the potential flow assumption. \\
The influence of the added mass in the system natural frequency \eqref{eq:f_nc} explains the frequency shift observed in figure \ref{fig:freq_disp_1} (b), and reported in figure \ref{fig:amass} (b) with a consistent normalisation. With a density ratio equal to $10.3$, \cite{khalak1997investigation} observed a similar behaviour in the normalised oscillation frequency of an elastically mounted cylinder without rotation, except that they found values slightly larger than $1.0$. We speculate that this difference might be mainly caused by the change in the sign of the angular velocity at the peak displacement and the associated change in vortex shedding timing. It is worth pointing out that the significance of the added mass effects clearly increases as the phase density ratio $\rho$ becomes smaller \citep{khalak1997investigation,khalak1999motions}. Further confirmation of the effectiveness of the coupling kinematics is provided by comparing the frequency ratio in the lock-in regime for different phase density values (see figure \ref{fig:amass} (c)). Within the lock-in $U_r$ interval the frequency ratio $f_y/f_{nc}$ is fairly insensitive to the phase density, as well as the lock-in threshold itself. On the contrary, oscillating cylinders with and without constrained rotation experience a significant variation of the lock-in width depending on the mass ratio \cite{vikestad2000added}. 
\section{Summary and outlooks} \label{sec:concl}
The present work introduces a model of water energy harvester belonging to the class of ALTs. Such technologies provide a new concept for the generation of clean and renewable energy from slow ocean/river currents, based on the amplification of the VIV phenomenon. Both numerical simulations and prototyping activities have shown that organised oscillating cylinders can result in higher conversion efficiency with respect to a wave energy harvester and more conventional devices, based on a normalised benchmarking \citep{bernitsas2008vivace}. \\
Our model consists of a transversely oscillating cylinder immersed in a free stream, with mechanically coupled rotation. This investigation provides a computational proof of concept which can potentially lead to an improvement of existing prototypes \citep{wang2016active,hobbs2012tree} for an augmented harnessed power, indeed. The fluid-structure system has been investigated by 2D FSI simulations based on DNS with IB forcing, at a Reynolds number equal to $100$. The impact of the mechanical coupling has been analysed over the parameter space spanned by reduced velocity and coupling radius. Furthermore, additional explorations with different phase density ratios have been carried out. \\
The kinematic coupling provides a new VIV scenario, in which the LAOs typical of the lock-in condition are magnified with respect to non-rotating case. Furthermore, the rotation--translation kinematic coupling broadens the reduced velocity domain where the lock-in condition takes place. These outcomes might potentially lead to innovative devices offering larger power outputs and extended optimal operating regions. Although the potential of rotation in enhancing the oscillation amplitude has been thoroughly investigated in previous studies, this represents the first concept device where no additional energy input is needed. With the suitable $(r,U_r)$ pair, the oscillatory rotation endows the cylinder with the suitable force distribution to achieve approximately a $360$ \% increment in oscillation amplitude with respect to non-rotating cylinders, on a peer hydrodynamic loading basis. It is worth pointing out that the coupling radius does not affect the frequency response for a given reduced velocity. The kinematic coupling inherently prevents the amplitude drop when increasing the reduced velocity, leading to a nearly monotonic growth of the peak oscillation amplitude and to the permanence of the locked state. Previous studies have shown that the galloping dynamics is characterised by an unbounded increase in oscillation amplitude, nevertheless, it does not involve a lock-in mechanism between structural response and vortex dynamics. Although galloping and lock-in dynamics may overlap, we define the condition associated to LAOs as lock-in, without the claim of a rigorous classification.  \\
The inspection of the wake pattern confirmed the wake-body synchronisation. The analysis of the wake-displacement frequency ratio reveals a clear tuning in most of the investigated cases, except that at large coupling radii/reduced velocity. Well-established patterns have been recognised and their relation with the oscillation dynamics has been elucidated. As a matter of fact, each pattern falls within a clearly delimited region of the frequency-amplitude domain. \\
Phase portraits revealed that cylinder response is generally periodic and it exhibits a strongly sinusoidal behaviour with null mean displacement. The phase difference between lift coefficient and cylinder displacement shed the light on the energy transfer between the fluid and the cylinder. For $U_r \leq 5.2$, the lift force is essentially in phase with the displacement, regardless of $U_r$ and $r$ values, whereas for any locked case, no phase jump is observed. The phase difference is found to gradually grow with the reduced velocity, with a maximum phase difference ($\approx 45^\circ$) inversely proportional to the coupling radius. Thus, the coupled rotation guarantees the phase alignment necessary to sustain LAO, making the oscillation amplitude grow indefinitely with the reduced velocity. This is inherently achieved by preventing the exact match between oscillation frequency and system natural frequency in vacuum. Such frequency offset is connected in turn with the added mass effect. Previous investigations correlate sufficiently large rotation rates with an increase of the added mass coefficient, which displaces the natural frequency of the system away from its oscillation frequency. In the present model the added mass coefficient is subject to limited changes with the reduced velocity, if compared with the uncoupled rotary motion case \citep{bourguet2014flow}, providing the necessary frequency shift to avoid the aforementioned phase jump. \\
As a concluding remark, we point out that the pursuit of a deterministic correlation behind the nature of such an extended lock-in region leads to a "circular" problem. This is an inherent consequence of the mutual dependence of reduced velocity, natural frequency of the system and added mass effect \citep{gabbai2005overview}. In view of this condition, fully coupled FSI simulations provide a suitable tool for investigating the problem. Thus, we speculate that the simple kinematic coupling $r\dot{\theta}=\dot{y}$ provides, for a wide collection of $(r,U_r)$ pairs, the balance of equivalent mass and peak rotation rate necessary to prevent the oscillator from the phase jump. \\
The present study offers several chances of generalization. In the first instance, the conceptualised model needs to be further investigated by including in the rigid body equation \eqref{eqn:dim_syst} the damping effect induced by the coupling with an electrical generator. Besides moving towards a more realistic scenario, the inclusion of a damping parameter can affect the peak oscillation amplitude for a wide range of phase density values \citep{skop1997new,govardhan2006defining}, therefore influencing the effectiveness of the system. Specifically, the amplitude of the structural response during lock-in and the band of reduced velocities over which the lock-in phenomenon exists is strongly dependent on the reduced damping parameter \citep{gabbai2005overview}. Since very large amplitude oscillations are achieved within the explored parameter space, the structural model can be further enhanced by considering a cubic stiffness term in equation \eqref{eqn:dim_syst} to account for stiffening effects under large displacement. \\
Although the low Reynolds number allowed for a simple elucidation of the FSI mechanism due to the lack of a chaotic dynamics, the effect of realistic flow condition, as well as the associated three dimensionality must be accounted for to get a full picture of the system operating performance. Eventually, a query for larger power outputs can be satisfied by coupling multiple cylinders in tandem/staggered configurations, once evaluated the optimal wake-cylinder interaction scenario \citep{kim2016performance}. \\
\backsection[Acknowledgements]{\\
This work has been partially supported by Italian Ministry of University and Scientific Research (MIUR) via PRIN2017 XFAST-SIMS Grant No. 20173C478N. The simulations were carried out on the high performance computing infrastructure of the Department of Mechanics Mathematics and Management, Polytechnic University of Bari (Italy).}
\backsection[Declaration of Interests]{\\
The authors report no conflict of interest.}
\appendix
\section{Constructive features}\label{sec:appB}
\begin{figure}
\centering
\includegraphics[width=0.8\columnwidth]{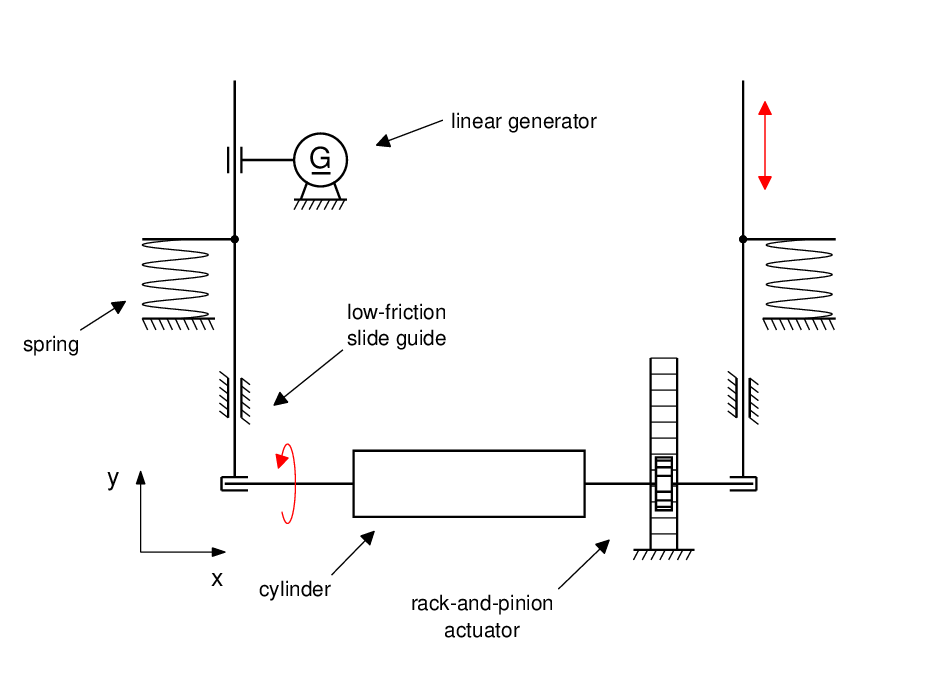}
\caption{Schematic of the proposed mechanical arrangement. The incoming flow takes place in the out-of-plane direction.}
\label{fig:mech_sck}
\end{figure}

In the present appendix we briefly address some practical aspects for the realisation of an energy harvester based on the FSI mechanism described in this manuscript. \\
The targeted energy harvester can be designed starting from the technical experience reported by \cite{bernitsas2006vivace}. The cylinder is constrained to oscillate in the cross-flow direction (corresponding to the y direction in figure \ref{fig:mech_sck}) by means of submerged rods sliding on low-friction guides. A recent contact-less magnetic slider for hydrokinetic energy harvesting has been patented \citep{bernitsas2021contact} with the aim of minimising friction and wear-related issues. A helical spring fixed on the static frame can provide the elastic force necessary to sustain large VIVs. A system inspired to our model must implement a shaft-mounted cylinder. Consequently, shaft and sliders must be connected by rolling bearings, in order to allow the rotation of the cylinder. We emphasise that the key feature of our model is the rotation-translation coupling. This can be realised by connecting the frame with the shaft through a rack-and-pinion mechanism. With the rack being joined to the device frame, the translating motion drives a rotation around the cylinder axis. In view of this solution, the model parameter $r$ actually represents the gear ratio. \\
We speculate that the energy conversion can be realised by means of a linear generator, such as that proposed by \cite{kim2017design} specifically for sea energy converters. This solution complies with any off-design operating condition in similar prototypes, since the oscillation amplitude is not constrained. This implies that, under lock-in conditions, the system can work with any inflow velocity while providing amplification of the limit-cycle oscillations. In this mechanical system, velocity-correlated damping effects might be attributed to the translation resistance offered by the generator and guides.
\section{Solver verification and convergence}\label{sec:appA}
\begin{figure}
\centering
\subfigure[]{\includegraphics[width=0.49\columnwidth]{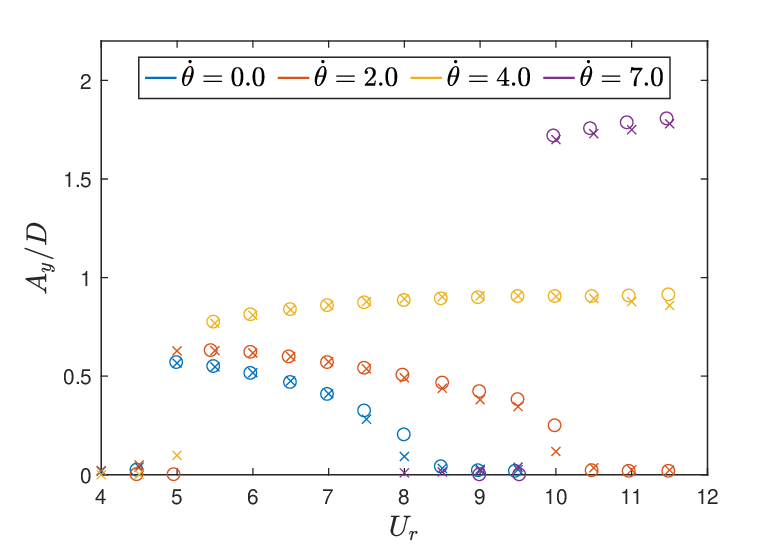}}
\subfigure[]{\includegraphics[width=0.49\columnwidth]{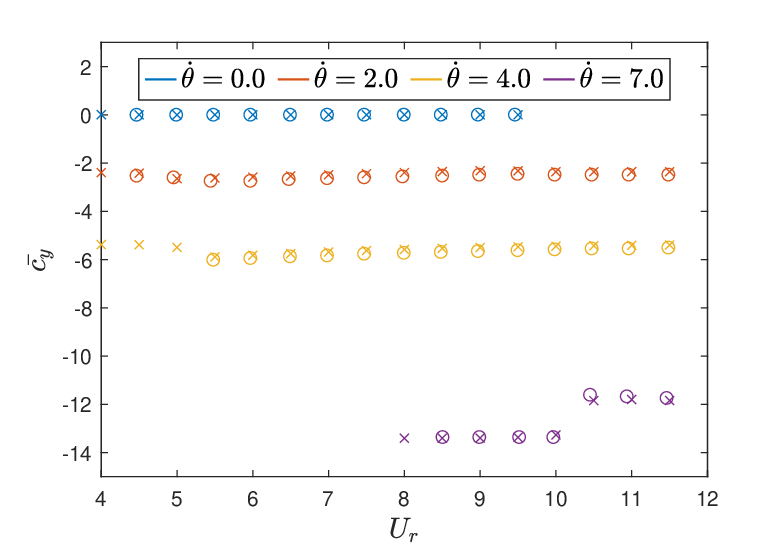}}
\caption{Comparison between the present numerical solution (crosses) and reference data from \cite{bourguet2014flow} for an elastically mounted cylinder with constrained rotation undergoing VIV (circles). The comparison is carried out in terms of oscillation amplitude (a) and time-averaged lift coefficient (b), for different rotation rates.}
\label{fig:valid3}
\end{figure}

\begin{figure}
\centering
\subfigure[]{\includegraphics[height=7.2cm]{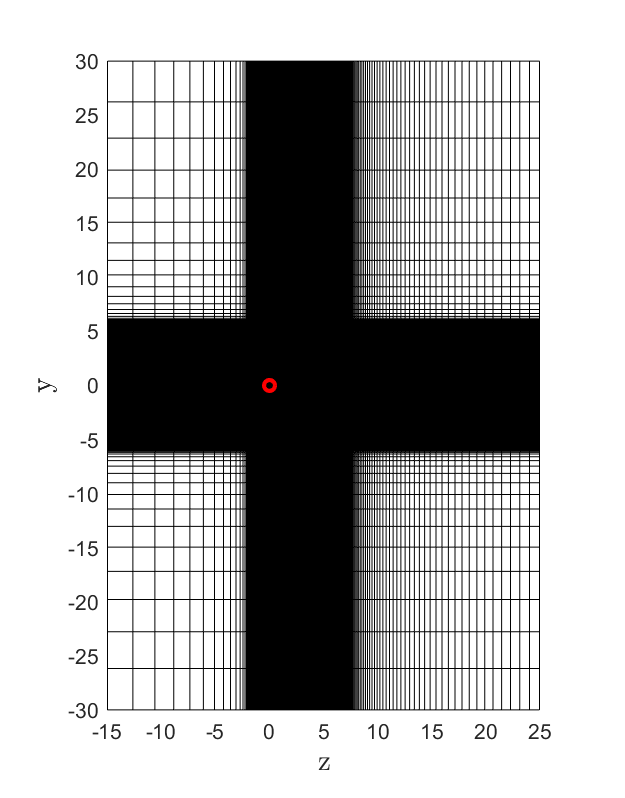}}
\subfigure[]{\includegraphics[height=7.2cm]{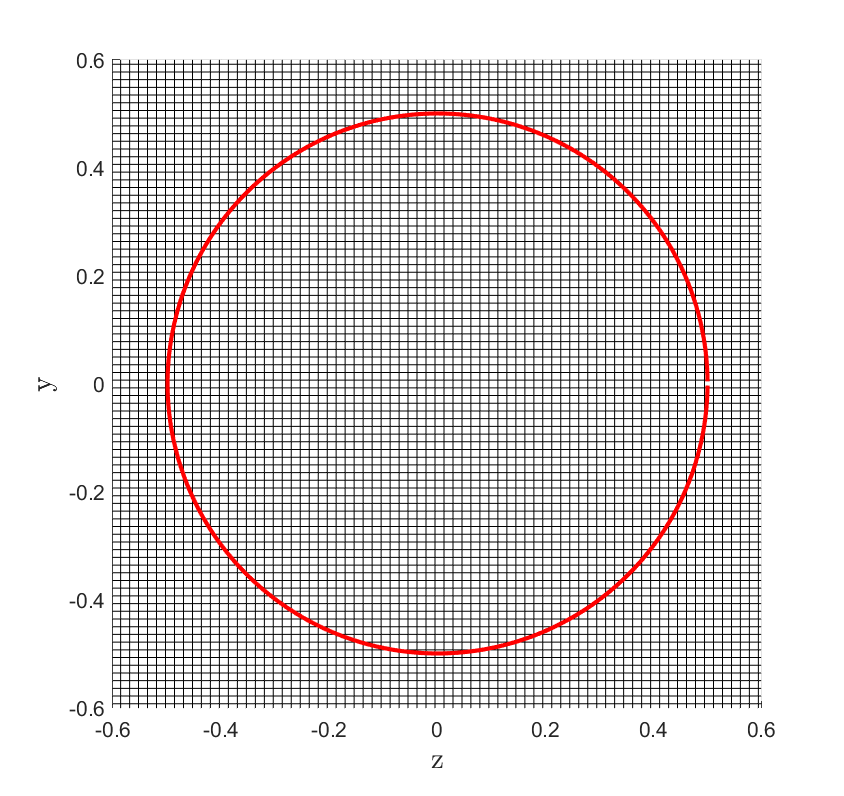}}
\caption{Computational domain and baseline grid. One line every four is plotted for the sake of clarity (a); Zoom-in of the circular cylinder immersed in the Cartesian grid (b)}
\label{fig:grid}
\end{figure}

We validate the tool employed in this computational campaign by comparing results from multiple test cases against data available in the literature. A broad validation has been already carried out in the presentation of the numerical tool \citep{de2016moving} against both two- and three-dimensional tests. Hence, all of the cases selected for this section are strictly related with the main focus of this work, both in terms of loading conditions and kinematic aspects. The following test are conducted using the baseline discretisation and the computational domain described in section \ref{sec:num_met}. \\
\begin{figure}
\centering
\subfigure[]{\includegraphics[width=0.49\columnwidth]{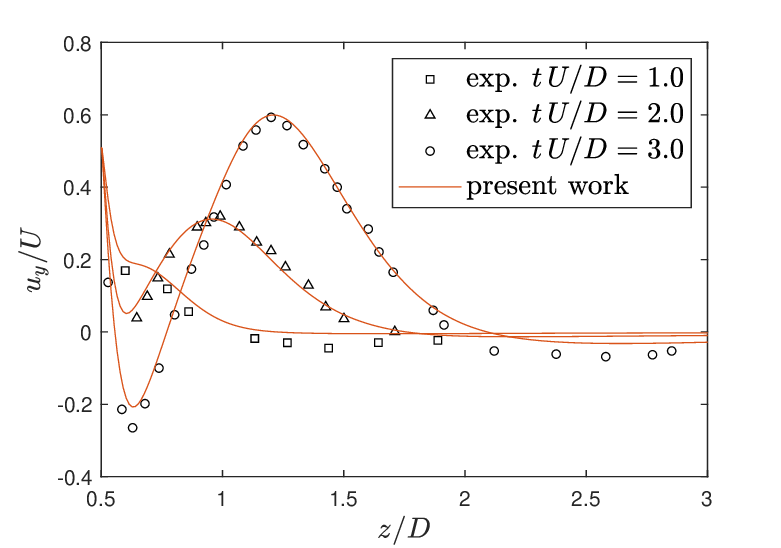}}
\subfigure[]{\includegraphics[width=0.49\columnwidth]{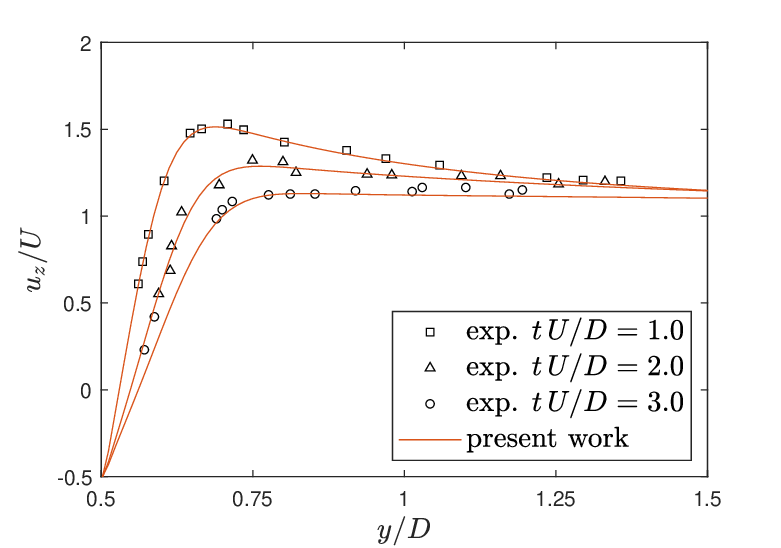}}
\caption{Spanwise (a) and streamwise (b) velocity profiles for an impulsively started rotating cylinder past a uniform flow at $Re=200$ and $\dot{\theta}=1.0$. Numerical results are compared with experimental data provided by \cite{coutanceau1985influence}.}
\label{fig:valid1}
\end{figure}
In the first instance we run the simulation of a rotating cylinder at $Re=200$ and rotation rate $\dot{\theta}=1.0$. The velocity profile of the spanwise and streamwise velocity components are compared with experimental data provided by \cite{coutanceau1985influence}. The former profile is taken over a horizontal symmetry line, whereas the latter over a vertical symmetry line. The cylinder and the fluid are initially at rest, and, given an impulsive fluid velocity and cylinder rotation rate, the velocity profiles are inspected at three subsequent time instants. Our numerical results match reasonably well the evolution of the boundary layer in the early stage of the flow observed in the experimental data (see figure \ref{fig:valid1}). This certifies that the no-slip condition is adequately enforced in the presence of a rotating interface and the related shear layers are consistently resolved. 
\begin{figure}
\centering
\includegraphics[width=0.6\columnwidth]{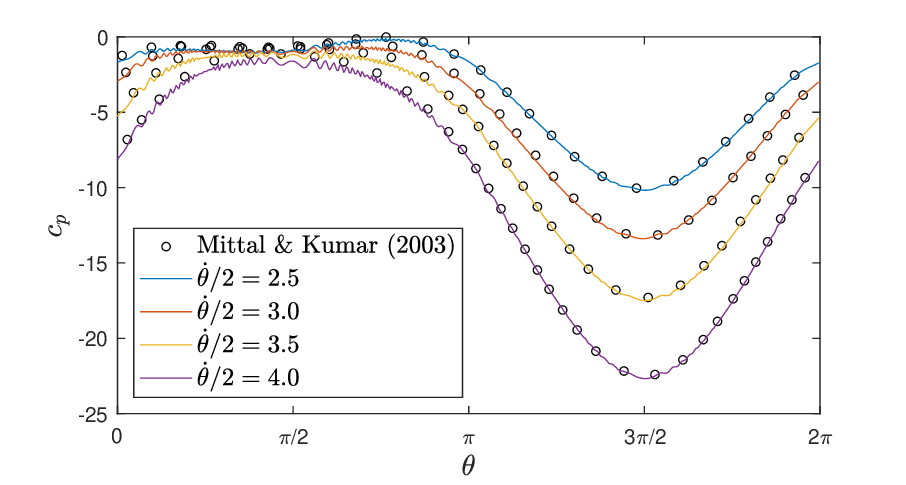}
\caption{Comparison of the pressure coefficient, defined as $c_p=2(p-p_0)/(\rho_f U^2)$, along the cylinder profile for a rotating cylinder at $Re=200$ with different rotation rates.}
\label{fig:valid2}
\end{figure}
The accuracy of the load computation technique has been likewise verified for a rotating cylinder, since this represents a potential source of uncertainty for the study problem. Four cases of a cylinder rotating at constant rotation rate and $Re=200$ have been simulated. Very large rotation rates have been considered, namely $\dot{\theta} D /(2U)=\{2.5,3.0,3.5,4.0 \}$, such that vortex shedding is suppressed. Figure \ref{fig:valid2} shows the pressure coefficient distribution over the cylinder profile, superposed with data from \cite{mittal2003flow}. The pressure coefficient is here defined as $c_p=2(p-p_0)/(\rho_f U^2)$, with $p_0$ being the pressure value at the inlet edge. Our computations have been able to accurately replicate the strongly asymmetric profile, which leads to an augmented lift force. \\
To check the accuracy of the FSI procedure, we replicated the results provided by \cite{bourguet2014flow} for an elastically mounted cylinder with forced rotation. The cylinder can only undergo cross-flow oscillations, which in turn are heavily affected by the rotation rate. This test resembles the hydrodynamic loading arising in the proposed system, therefore, it is taken as a final validation step. All cases are characterised by $Re=100$, $\rho=10$, whereas the rotation rates comprise the values $\{0.0,2.0,4.0\}$. The maximum rotation ratio is close to the largest value observed in our investigation. The reduced velocity is spanned in between $U_r=4.0$ and $U_r=12.0$ with $\Delta U_r=0.5$ increments. The oscillation amplitude and the time-averaged lift coefficient are compared with the numerical results by \cite{bourguet2014flow} in figure \ref{fig:valid3}. Both the maximum oscillation amplitude and the width of the lock-in region agree well with those reported in the reference study. The comparison is carried out up to a rotation rate equal to $\dot{\theta}=7.0$, which is fairly close to the peak rotation rate encountered in this study. Furthermore, a few cases have been run at $\dot{\theta}=8.0$, confirming the suppression of large amplitude vibrations. Analogously, the dependence of the amplitude on the rotation rate is well reproduced. Similar to that of a stationary cylinder, the magnitude of the negative lift coefficient increases with the increase of the rotation rate of the cylinder. \\ 
A grid convergence study was performed on the system under investigation with $U_r=8.0$ and $r=0.4$, upon the occurrence of LAOs. The baseline grid tested consists of $801 \times 801$ grid nodes, with uniform grid spacing in the region $[-2D,8D]\times[-6D,6D]$, corresponding to a local resolution of approximately 60 grid nodes along the cylinder diameter. This grid is compared with those with $581 \times 581$ and $1081 \times 1081$, which corresponds to $40$ and $80$ grid nodes along the diameter, respectively. A sketch of the Cartesian grid and domain size is provided in figure \ref{fig:grid}. These grids are $0.53$ and $1.82$ the size of the baseline grid in terms of total node count. All simulations are performed by keeping a constant refinement ratio between Lagrangian markers and local Eulerian grid size as $\Delta l/\Delta x=0.5$, therefore $252$, $378$ and $504$ markers are employed, respectively. In order to assess the convergence of both spatial and temporal discretization schemes this comparison is conducted at constant $\text{CFL}=0.2$, leading to different time-step size. 
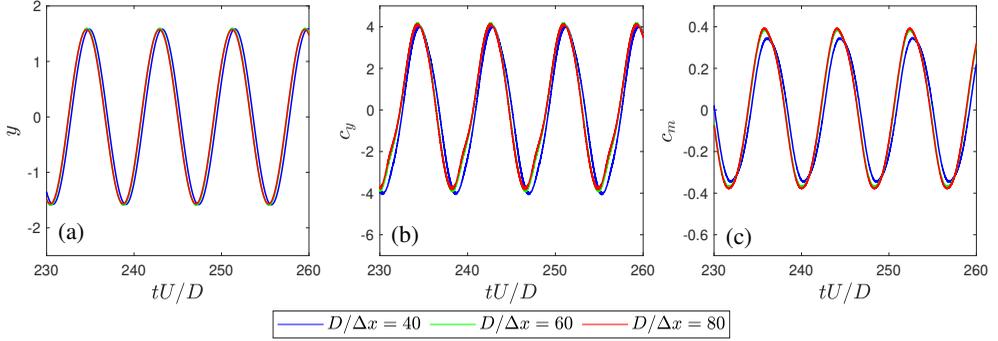
\begin{figure}
\begin{tikzpicture}
      \node[anchor=south west,inner sep=0] (a) at (0,-0.1) {\includegraphics[width=0.99\columnwidth]{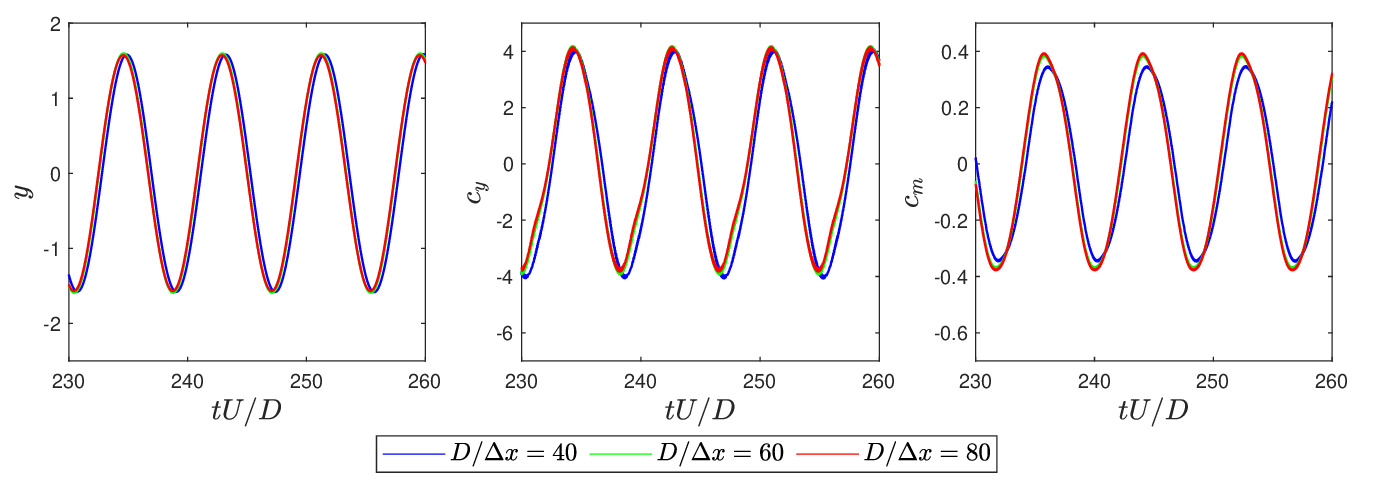}};
      \begin{scope}[x={(a.south east)},y={(a.north west)}]
      \node [align=center] at (0.075,0.3) {(a)};
      \node [align=center] at (0.405,0.3) {(b)};
      \node [align=center] at (0.735,0.3) {(c)};
      \end{scope}
\end{tikzpicture}
\caption{Time traces of (a) vertical displacement, (b) lift coefficient and (c) moment coefficient for flow-induced oscillations at $U_r=8.0$ using the three discretization sets tested.}
\label{fig:conv}
\end{figure}
\begin{table}
\centering
\begin{tabular}{ccccc}
\quad & $(y/D)^{\text{max}}$ (\%) & $T_y U/D$ (\%) & $c_y^{\text{max}}$ (\%) & $T_{c_y} U/D$ (\%) \\
coarse & 1.1 & 0.10 & 3.2 & 0.23   \\
baseline & 0.11 & 0.0061 & 0.42 & 0.017  \\
\end{tabular}
\caption{Percentage of error with regard to the fine mesh for some of the parameters monitored in the convergence analysis.}
\label{tab:conv}
\end{table}
The comparison is presented in terms of percentage variation of maximum value and period (computed as a root mean square over the last 20 cycles) of the vertical displacement $y/D$ and lift coefficient $c_y$. Data are provided schematically in table \ref{tab:conv}, whereas time traces of $y/D$, $c_y$ and $c_m$ are plotted in figure \ref{fig:conv} for the three grids examined. All parameters show a much lower variation from the baseline grid to the fine grid than from the coarse one. Thus, the baseline grid can be considered within the asymptotic range in terms of spatial and temporal discretization. The grid convergence above illustrated, although performed for a single case, is expected to be a demonstration of grid convergence for the system dynamics investigated in the present work. Furthermore, we have tested different cases over the same grids, which provided similar root-mean-square variations as those mentioned above. These results are not shown here for the sake of brevity. \\
In the present study the size of the computational domain was selected relying on the results provided by \citep{zhao2014vortex,kang1999laminar}, in which a comparison of different domain sizes was performed with identical boundary conditions. We selected the extension $[-15D,25D] \times [-30D,30D]$ since it provides minimal blockage effect with a limited computational expense on a similar rotation-rate basis.

\bibliography{jfm}
\bibliographystyle{jfm}

\end{document}